\documentclass[aps, floatfix, a4paper, nofootinbib,
superscriptaddress, 11pt]{revtex4}

\usepackage{graphicx,float,tikz}
\usepackage[all]{xy}
\usepackage{amsmath,upgreek}
\usepackage{amsfonts}
\usepackage{amsthm}   
\usepackage{bbold}
\usepackage{bm}
\usepackage{amssymb}
\usepackage{color}
\usepackage{epsfig}		
\usepackage{graphicx,epstopdf}
\usepackage{subfigure}
\usepackage{pdfpages}
\usepackage{dsfont}
\usepackage{appendix}
\usepackage[colorlinks,hyperindex]{hyperref}

\definecolor{green1}{RGB}{0,128,0} 
\hypersetup{backref=true,pagebackref=true}
\hypersetup{%
  colorlinks = true,
  linkcolor  = blue,
  citecolor = red,
}
\usepackage{bookmark,textgreek}
\usepackage{hyperref,color,xcolor}
\hypersetup{hidelinks,hyperindex=true,colorlinks=true,breaklinks=true,urlcolor=red}
\hypersetup{%
  colorlinks = true,
  linkcolor  = blue
}

\begin{document}

\title{Magnetic-field-driven topological phase transition in the holograpic Weyl semimetal}

\author{R.C.L. Bruni}
\email{bruni.r.c.l@gmail.com}
\affiliation{The University of Queensland, School of Mathematics and Physics,\\
Queensland 4072, Australia.}
\author{Luiz F. Ferreira}
\email{lffaulhaber@gmail.com}
\affiliation{Instituto de Física y Astronomia, Universidad de Valparaiso, \\ A. Gran Bretana 1111, Valparaiso, Chile}
\author{Diego M. Rodrigues}
\email{diegomhrod@gmail.com}
\affiliation{Instituto de Física Teórica, Universidade Estadual Paulista,\\ R. Dr. Bento T. Ferraz 271, Bl. II, São Paulo 01140-070, SP, Brazil}

\begin{abstract}

We study the magnetic field effects on the quantum critical point (QCP) in the holographic Weyl semimetal model. We show that it increases quadratically with the magnetic field for weak field and linear with the magnetic field for strong field. Our findings are compatible with previous results in the literature from other approaches. 

\end{abstract}


\maketitle

\section{Introduction}
The investigation about the {\it topological states of matter}, {\it i.e.} materials that do not fit in the usual Ginzburg-Landau criteria being classified by topology rather than symmetry (or breaking of it) has become increasingly stronger in condensed matter physics. In such materials the momentum-space topology of the Fermionic ground states, most commonly manifested in their electronic band structure, plays a fundamental role \cite{schnyder_classification_2008,chiu_classification_2016}. The physical realization of these states includes topological insulators (TI) \cite{kane_z_2_2005,fu_topological_2007,Hasan_colloquim_2010}, topological superconductors (TSC) \cite{qi_topological_2011,kallin_chiral_2016,sato_topological_2017} and Weyl semi-metals (WSM) \cite{armitage_weyl_2018,hosur_recent_2013,jia_weyl_2016,wehling_dirac_2014,gao_topological_2019,vafek_dirac_2014}.

The low energy effective field theory description of WSM's is a linear gapless effective Hamiltonian resembling a relativistic {\it Weyl fermion} from particle physics literature \cite{weyl_elektron_1929,herring_accidental_1937}, whose geometric {\it locus} in momentum space defines a cone centered at given value of momenta and energy, the {\it Weyl cone} \cite{burkov_topological_2011,halasz_time-reversal_2012,delplace_topological_2012,pesin_mott_2010,wan_topological_2011,yang_quantum_2011,shuichi_murakami_phase_2007,hosur_charge_2012,burkov_weyl_2011-1,zyuzin_weyl_2012,zyuzin_topological_2012,goswami_axionic_2013,grushin_consequences_2012,vazifeh_electromagnetic_2013}. These cones must always occur in an even number throughout the \textbf{k}-space, as a result of the {\it Fermion doubling theorem} \cite{nielsen_absence_1981, nielsen_absence_1981(2)}. This last feature, together with their associated topological charge \cite{volovik_quantum_2007,klinkhamer_emergent_2005}, endows the Weyl cones with a topological stability against perturbations that preserves the original symmetries of the system and leads to a series of interesting physical consequences, \textit{e.g.}, Fermi-Arcs \cite{xu_discovery_2015,lu_experimental_2015,belopolski_criteria_2016,liu_chiral_2013,fukushima_chiral_2008,son_chiral_2013} and the Chiral magnetic effect \cite{kharzeev_chiral_2009,huang_observation_2015,zhang_signatures_2016}.

The development of the AdS/CFT correspondence or, more generally known as the gauge/gravity duality, over the past years has become an important tool for studying strongly-coupled systems. Originally, the AdS/CFT correspondence states a duality between a gravity theory in a ($ d+1 $)-dimensional anti-de Sitter (AdS) space and $\mathcal{N}=4$ super-Yang-Mills theory, living on d-dimensional boundary of AdS space. Namely, the physics of the strongly-coupled theory on $ d $-dimensional boundary of AdS space is encoded in the weakly-coupled classical gravity living on a ($ d+1 $)-dimensional AdS space. Nowadays, the extension of this duality has allowed to model a variety of condensed matter systems \cite{Zaanen:2015oix,hartnoll_holographic_2018,Nastase:2017cxp}. In particular, a holographic Weyl-semimetal model and its quantum phase transition to a trivial semimetal, has been constructed in \cite{Landsteiner:2015lsa,Landsteiner:2015pdh} by including  deformations in the gravitational theory which mimic the essential features of a Weyl semimetal effective field theory.

At zero temperature, the holographic Weyl semimetal  \cite{Landsteiner:2015pdh} exhibits a topological quantum phase transition from Weyl semimetal to a topological trivial phase, where the order parameter is the anomalous Hall conductivity. The quantum phase transition occurs at a critical value of the effective coupling $(M/b)_{\textrm{crit}}$, where $M$ and $b$ are parameters included, via proper boundary conditions, in the holographic model and which are dual to the mass deformation and the axial gauge field in the Weyl semimetal effective field theory. Thus, due to the inclusion of essential features, this approach provides a great tool to investigate various properties of Weyl semimetal physics, such as  surface state \cite{Ammon:2016mwa}, odd  viscosity \cite{Landsteiner:2016stv}, conductivity \cite{Grignani:2016wyz}, axial Hall conductivity \cite{Copetti:2016ewq}, topological invariants \cite{Liu:2018djq}, momentum relaxation \cite{Zhao:2021qfo,Zhao:2021pih}, nodal line semimetal \cite{Liu:2018bye,Liu:2020ymx,Rodgers:2021azg} and Weyl-Z$_{2}$ semimetal \cite{Ji:2021aan}. Besides, see \cite{Landsteiner:2019kxb} for a recent review. Further applications of holography in Weyl semimetal physics can be found for instance in \cite{Gursoy:2012ie,Jacobs:2015fiv,Hashimoto:2016ize,Liu:2018spp,Song:2019asj,Ji:2019pxx,Baggioli:2020cld,Juricic:2020sgg,BitaghsirFadafan:2020lkh,Gao:2023zbd}.

Motivated by recent experimental progress on the role played by the magnetic field as a driver of topological phase transitions for instance in \emph{Fe}-doped heterostructures \cite{Shiogai_et_al} and in a trivial semimetal \cite{Mayo_et_al}, we extend the previous framework of the holographic Weyl semimetal model \cite{Landsteiner:2015lsa} and its quantum phase transition to a trivial semimetal \cite{Landsteiner:2015pdh} by including a finite magnetic field in order to study the behavior of the quantum critical point as a function of the magnetic field. The paper is organized as follows: in Sec. \ref{sec2} we revisit the Landau level picture in the effective field theory description of Weyl Semimetals. In Sec. \ref{sec3} we present the holographic model for the Weyl semimetal in a magnetic field and our numerical results. Finally, in Sec. \ref{sec4} we present our conclusions and discussions.

\section{Revisiting the Effective Description of Weyl Semimetals in a Magnetic Field: Landau Levels}
\label{sec2}

In this section we revisit the low-energy effective description of a Weyl Semimetal in the presence of a uniform magnetic field and compute the energy spectrum associated with the Landau Levels.

Weyl semimetals are exotic materials whose band structure shows points around which the low-energy excitations are a pair of left- and right-handed Weyl fermions, which are chiral massless relativistic particles, lying exactly at the Fermi level (for a short review see \cite{https://doi.org/10.48550/arxiv.1603.02821}). Their low-energy effective field theory description can be captured by a Lorentz-breaking Dirac lagrangian \cite{Colladay:1998fq} coupled to an external electromagnetic potential $A_{\mu}$ given by
\begin{equation}\label{diraclagrangian}
	\mathcal{L} = \bar{\Psi}\left(\gamma^{\mu}\left(i\,\partial_{\mu} - e\,A_{\mu}\right) - \gamma^5\, \bm{\gamma}.\textbf{b} - M \right)\Psi, 
\end{equation}
where $ \gamma^{\mu} = (\gamma^0, \bm{\gamma}) $ are the gamma matrices which satisfy the Clifford algebra $ \left\lbrace \gamma^{\mu},\gamma^{\nu} \right\rbrace = 2\,\eta^{\mu\nu}\mathds{1}_4 $, where $\eta_{\mu\nu}$ is the Minkowski metric whose signature we take here to be $ (+,-,-,-) $. $\gamma^5$ is the chiral matrix defined as $\gamma^5 = i\,\gamma^{0}\,\gamma^{1}\,\gamma^{2}\,\gamma^{3}$, and the electromagnetic potential is generally given by $A_{\mu} = (\phi,-\textbf{A})$ where $\phi$ is the eletric potential and $\textbf{A}$ is the vector potential. The vector $ \textbf{b} $ is an axial vector responsible for breaking Lorentz symmetry, and which we will take it in the $ \hat{z} $ direction, i.e, $ \textbf{b} = b\,\hat{z} $, and will be the separation in momentum between the Weyl nodes. Finally, $M$ is the mass deformation.

The Dirac equation coming from the variation of the above lagrangian with respect to $\bar{\Psi}$ reads
\begin{equation} \label{diraceq}
	\left(\gamma^{\mu}\left(i\,\partial_{\mu} - e\,A_{\mu}\right) - \gamma^5\, \bm{\gamma}.\textbf{b} - M \right)\Psi = 0.
\end{equation}
Since in this section we are only interested in study the system in a constant and uniform magnetic field in the $\hat{z}$ direction, we take the electromagnetic potential to be $ A_{\mu} = (0, - \textbf{A}) $, where the vector $ \textbf{A} $ will be chosen in the symmetric gauge, which is given by $ \textbf{A} =(-\frac{B\,y}{2},\frac{B\,x}{2},0) $, where B is the magnetic field. One can clearly see that this choice for the vector potential yields $\nabla\times \textbf{A} = B\,\hat{z}$. Thus, the Dirac equation \eqref{diraceq} takes the form
\begin{equation} \label{diraceq2}
	\left(\gamma^{\mu}(i\,\partial_{\mu}) + e\,\bm{\gamma}.\textbf{A} - \gamma^5\, \bm{\gamma}.\textbf{b} - M \right)\Psi = 0.
\end{equation}
Now, going to momentum space (using the correspondence principle $i\partial_{\mu} = p_{\mu} = (E,-\textbf{p})$) we have
\begin{equation} \label{diraceq3}
	\left(\gamma^{0}\,E+\bm{\gamma}.\left(-\textbf{p} + e\,\textbf{A}\right) - \gamma^5\, \bm{\gamma}.\textbf{b} - M \right)\Psi = 0,
\end{equation}
which can be written as
\begin{equation} \label{diraceq4}
	\left(\bm{\gamma}.\left(\textbf{p} - e\,\textbf{A}\right) + \gamma^5\, \bm{\gamma}.\textbf{b} + M \right)\Psi = \gamma^{0}\,E\,\Psi.
\end{equation}
Multiplying both sides from the left by $\gamma^{0}$ and, using that $(\gamma^{0})^2 = \mathds{1}$, we can turn the Dirac equation \eqref{diraceq} into a eigenvalue problem.
\begin{equation}
	\hat{H}\Psi = E\Psi,
\end{equation}
where the Hamiltonian operator $ \hat{H} $ is given by:
\begin{equation}
	\hat{H} = \left(\gamma^{0}\,\bm{\gamma}.\left(\textbf{p} - e\,\textbf{A}\right) + b\,\gamma^{0}\,\gamma^5\,\gamma^{3} + \gamma^{0}\,M \right),
\end{equation}
where we used that $\textbf{b}=b\,\hat{z}$. For our particular case, it will be convenient to work in the Weyl representation for the gamma matrices, which is given by:
\begin{equation}
	\gamma^{0} = \left(\begin{matrix} 
		0 & \mathds{1}_2\\
		\mathds{1}_2 & 0\\
	\end{matrix} \right), \quad \gamma^{i} = \left(\begin{matrix} 
		0 & \sigma^{i}\\
		-\sigma^{i} & 0\\
	\end{matrix} \right), \quad \gamma^{5} = \left(\begin{matrix} 
		-\mathds{1}_{2} & 0\\
		0 & \mathds{1}_{2}\\
	\end{matrix} \right), 
\end{equation}
where $ \mathds{1}_2 $ is the $2\times2$ identity matrix, and $\sigma_{i}$ are the Pauli matrices, given by:
\begin{equation}
	\sigma_{x} = \left(\begin{matrix} 
		0 & 1\\
		1 & 0\\
	\end{matrix} \right), \quad \sigma_{y} = \left(\begin{matrix} 
		0 & -i\\
		i & 0\\
	\end{matrix} \right), \quad \sigma_{z} = \left(\begin{matrix} 
		1 & 0\\
		0 & -1\\
	\end{matrix} \right). 
\end{equation}
Using the gamma matrices defined above, the Hamiltonian operator takes the form
\begin{equation}\label{HamiltonianOp}
	\hat{H} = \left(\begin{matrix} 
		\bm{\sigma}.\left(\textbf{p} - e\,\textbf{A}\right)+b\,\sigma_{z} & M\,\mathds{1}_2\\
		M\,\mathds{1}_2 & -\bm{\sigma}.\left(\textbf{p} - e\,\textbf{A}\right)+b\,\sigma_{z}\\
	\end{matrix} \right).
\end{equation}
Notice that the system have a well defined \textit{helicity} given by the eigenvalues of the operator $\hat{h} = \hat{\bm{\sigma}} \cdot \big( \hat{ \overline{\textbf{p}}}  - e \hat{A}\big)$ \footnote{ Usually the helicity operator is defined to be  $\hat{ h }  = \hat{\bm{\sigma}} \cdot  \hat{\textbf{p}} $, which is generalized to this definition according to the usual minimal coupling prescription.} - with $ \hat{ \overline{\textbf{p}}} $ being the momenta shifted by $ \pm \textbf{b} = b \hat{z}$ which coincides with the chirality in the massless limit, even though it is well known that in the absence of external electromagnetic field, it is also possible to define an effective chirality operator if $ b > M$ \cite{goswami_axionic_2013}.

Now, in order to proceed with the computation, let us take a look on the term $ \hat{\bm{\sigma}}.\left(\hat{\textbf{p}} - e\,\hat{\textbf{A}}\right)+b\,\hat{\sigma}_{z} $. In components and, using the symmetric gauge for the vector potential $ \hat{\textbf{A}} $, we have
\begin{equation}\label{quantum1}
	\hat{\sigma}_{x}\,\hat{\Pi}_{x} + \hat{\sigma}_{y}\,\hat{\Pi}_{y} + \hat{\sigma}_{z}\,\left(\hat{p}_{z} + b\,\mathds{1}\right),
\end{equation}  
where we introduced the operators $\hat{\Pi}_{x}$ and $\hat{\Pi}_{y}$, defined by
\begin{equation}
	\hat{\Pi}_{x}  := \hat{p}_x + \dfrac{ eB\, \hat{y} }{ 2}; \quad \hat{\Pi}_{y}  := \hat{p}_y - \dfrac{eB\, \hat{x}}{2},
\end{equation}
which satisfy the canonical commutation relation $ [\hat{\Pi}_{x}, \hat{\Pi}_{y}] = i\,eB$. Now, in terms of the ``$\Pi$'s" operators above defined, we introduce the annihilation and creation operators, $\hat{a}$ and $\hat{a}^{\dagger}$ \cite{PhysRevB.83.195413,RevModPhys.81.109,PhysRevB.87.245131}:
\begin{equation}
	\hat{a} := \sqrt{\dfrac{1}{2\, eB}} \left(\hat{\Pi}_{x} + i\, \hat{\Pi}_{y}\right); \quad \hat{a}^\dagger := \sqrt{\dfrac{1}{2\,eB} } \left(\hat{\Pi}_{x} - i\,\hat{\Pi}_{y} \right)  \, , \label{ladder}
\end{equation}
which satisfy the canonical commutation relation $ [\hat{a}, \hat{a}^{\dagger }] = 1 $. Thus, eq \eqref{quantum1} can be written as
\begin{equation}
	\sqrt{2\,eB}\left(\hat{S}_{-}\,\hat{a} + \hat{S}_{+}\,\hat{a}^{\dagger}\right)+2\,\hat{S}_{z}\left(b + \hat{p}_z\right)\, ,
\end{equation}
where $\hat{S}_{\pm} = \frac{1}{2}\,\hat{\sigma}_{\pm}:=\frac{1}{2}\, \left(\hat{\sigma}_{x} \pm i\,\hat{\sigma}_{y}\right)$ and $\hat{S}_z:=\frac{1}{2}\,\hat{\sigma}_{z}$ are the spin operators. Therefore, in terms of these definitions, the hamiltonian operator, given by eq. \eqref{HamiltonianOp}, becomes
\begin{equation}\label{HamiltonianOp2}
	\hat{H} = \left(\begin{matrix} 
		\sqrt{2\,eB}\left(\hat{S}_{-}\,\hat{a} + \hat{S}_{+}\,\hat{a}^{\dagger} \right) + 2\,\hat{S}_{z}\,\hat{p}_z + 2\,b\,\hat{S}_{z} & M\,\hat{\mathds{1}}_2\\
		M\,\hat{\mathds{1}}_2 & -\left(\sqrt{2\,eB}\left(\hat{S}_{-}\,\hat{a} + \hat{S}_{+}\,\hat{a}^{\dagger} \right) + 2\,\hat{S}_{z}\,\hat{p}_z\right) + 2\,b\,\hat{S}_{z} \\
	\end{matrix}\right).
\end{equation}

\begin{figure}[H]
	\centering
	\includegraphics[scale=0.7]{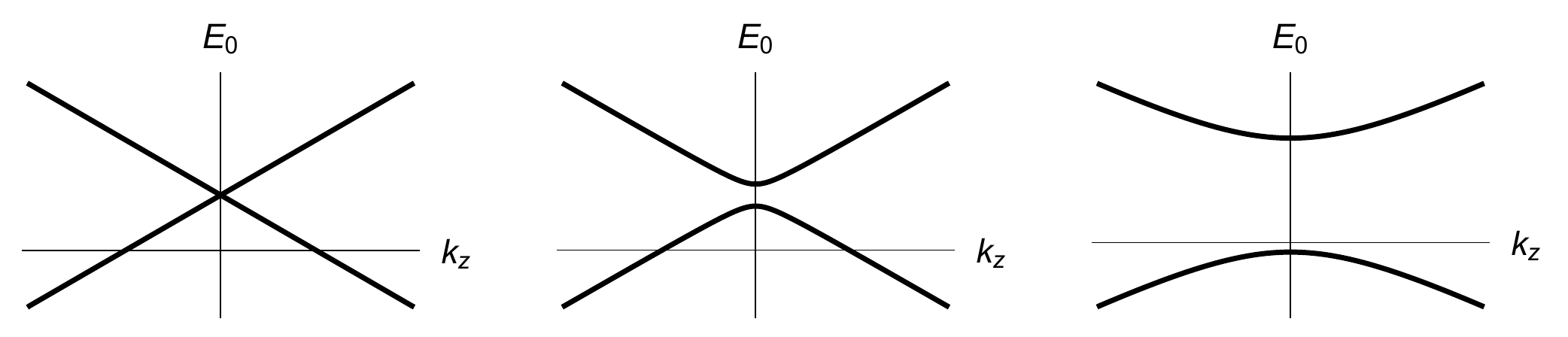}
	\caption{Energy dispersion for the lowest Landau level (LLL), $E_0$, in three different scenarios: (i) $M=0$; (ii) $\frac{M}{b}<1$ and (iii) $\frac{M}{b}>1$. Note that between $\frac{M}{b}<1$ and $\frac{M}{b}>1$ a quantum phase transition takes place and, for $\frac{M}{b}>1$, the system is gapped and become a trivial semimetal, i.e., an insulator.}
	\label{fig:ll}
\end{figure}

Now, we have to define the ground state, which is going to correspond to the lowest Landau level (LLL). It is characterized by the tensor product\footnote{Since the spin does not depend on the spatial degrees of freedom, the operator $\hat{S}$ commutes with all the spatial operators. In particular, it commutes with the momentum operator $\hat{p}$, i.e, $ \left[S_i, p_{j} \right] = 0 $. Therefore, it acts only on the spin part $\left|\frac{1}{2}, +\frac{1}{2}\right\rangle$ and leave the rest unchanged.}
\begin{equation*}
	\left|0,k_z \right\rangle \otimes \left|\dfrac{1}{2},+\dfrac{1}{2}\right\rangle := \left|0,k_z, +\dfrac{1}{2}\right\rangle,
\end{equation*}
where $\left|\frac{1}{2},+\frac{1}{2}\right\rangle$ is the eigenstate of the spin-1/2 operator $S_z$ with eingenvalue $+\frac{1}{2}$\footnote{In this case, the spin up with $m_s = \frac{1}{2}$ is the ground state because the spining motion gives rise to a magnetic dipole moment given by $\vec{\mu}_{S} = \frac{e}{2\,mc}\,\vec{S}$ which, when interacting with the external magnetic field $\vec{B}$, in turn generates a coupling term $-\vec{\mu}_{S}.\vec{B}$, which is minimized when the spin is aligned with the magnetic field.}. Note that the term $(\sqrt{2\,eB}(\hat{S}_{-}\,\hat{a} + \hat{S}_{+}\,\hat{a}^{\dagger}) + 2\,\hat{S}_{z}\,\hat{p}_z + 2\,b\,\hat{S}_{z})$ in \eqref{HamiltonianOp2} is annihilated by the ground state $\left|0,k_z, +\frac{1}{2}\right\rangle $ since $\hat{S}_{-}\,\hat{a}\left|0,k_z, +\frac{1}{2}\right\rangle=0$ and $\hat{S}_{+}\,\hat{a}^{\dagger}\left|0,k_z, +\frac{1}{2}\right\rangle=0$. Thus, the matrix elements of \eqref{HamiltonianOp2}, $\left\langle +\frac{1}{2},k_z,0\right|\hat{H}\left|0,k_z, +\frac{1}{2}\right\rangle$, are given by
\begin{equation}\label{MatrixElemGS}
	H = \left(\begin{matrix} 
		k_z + b & M\\
		M & -(k_z - b) \\
	\end{matrix}\right),
\end{equation}
where we used that $\hat{p}_{z}\,\left|0,k_z, +\frac{1}{2}\right\rangle = k_{z}\,\left|0,k_z, +\frac{1}{2}\right\rangle$ and $\hat{S}_{z}\,\left|0,k_z, +\frac{1}{2}\right\rangle = +\frac{1}{2}\,\left|0,k_z, +\frac{1}{2}\right\rangle$. Thus, diagonalizing this matrix we find that the ground state is linearly dispersing (see Fig.\ref{fig:ll}) and given by
\begin{equation}\label{energy_LLL}
	E = b\pm\sqrt{k_z^2 + M^2}; \quad n=0.
\end{equation}
\noindent
Note that the energy dispersion for the LLL \eqref{energy_LLL} is independent of the magnetic field $ B $ and purely chiral, i.e., right (left) movers for the negative (positive) chirality node \cite{PhysRevResearch.2.033511}. 

For the excited Landau levels, $n\geq1$, the most general eigenstate is characterized by $\left|n,k_z,\pm\frac{1}{2}\right\rangle$. Therefore, in order to construct the matrix elements, it will be convenient to square the hamiltonian operator \eqref{HamiltonianOp2} and writing them as $\left\langle \pm\frac{1}{2},k_z,n\right|\hat{H}^2\left|n,k_z, \pm\frac{1}{2}\right\rangle$. By doing this computation we find that the hamiltonian operator \eqref{HamiltonianOp2} squared is given by
\begin{equation}\label{HamiltonianSquared}
	\hat{H}^2 = \left(\begin{matrix} 
		2\,eB\left(\hat{a}^{\dagger}\,\hat{a} + \frac{1}{2} - \hat{S}_z\right) + (\hat{p}_z + b)^2 + M^2 & 4\,Mb\,\hat{S}_z\\
		4\,Mb\,\hat{S}_z & 2\,eB\left(\hat{a}^{\dagger}\,\hat{a} + \frac{1}{2} - \hat{S}_z\right) + (\hat{p}_z - b)^2 + M^2 \\
	\end{matrix}\right).
\end{equation}
Now, acting with this operator on the spin-up sector, with matrix elements given by $\left\langle +\frac{1}{2},k_z,n\right|\hat{H}^2\left|n,k_z, +\frac{1}{2}\right\rangle$, we find
\begin{equation}
	H^2_{\uparrow} = \left(\begin{matrix} 
		2\,eB\,n + (k_z + b)^2 + M^2 & 2\,Mb\\
		2\,Mb & 2\,eB\,n + (k_z - b)^2 + M^2 \\
	\end{matrix}\right).
\end{equation}
Analogously, for the spin-down sector, with matrix elements given by $\left\langle -\frac{1}{2},k_z,n\right|\hat{H}^2\left|n,k_z, -\frac{1}{2}\right\rangle$, we obtain
\begin{equation}
	H^2_{\downarrow} = \left(\begin{matrix} 
		2\,eB\,(n+1) + (k_z + b)^2 + M^2 & -2\,Mb\\
		-2\,Mb & 2\,eB\,(n+1) + (k_z - b)^2 + M^2 \\
	\end{matrix}\right).
\end{equation}
Therefore, the excited Landau levels ($n\geq1$) for both spin up and down sectors, respectively, are given by
\begin{eqnarray}
	E_{n}^{\uparrow} &=& \pm \sqrt{E_{0}^2+\beta\,n}; \quad n\geq1, \label{Energyup}\\
	E_{n}^{\downarrow} &=& \pm \sqrt{(E_{n}^{\uparrow})^2+\beta}, \label{Energydown}
\end{eqnarray}
where $E_0 = b\pm\sqrt{k_z^2+M^2}$ is the LLL, and $\beta = 2\,eB$. From above equations one can see that there is an energy gap $\Delta E_{n}^2$ between the up and down spins, given by
\begin{equation}
	\Delta E_{n}^2 := (E_{n}^{\downarrow})^2 - (E_{n}^{\uparrow})^2 = \beta = 2\,eB.
\end{equation}

\begin{figure}[H]
	\centering
	\includegraphics[scale=0.7]{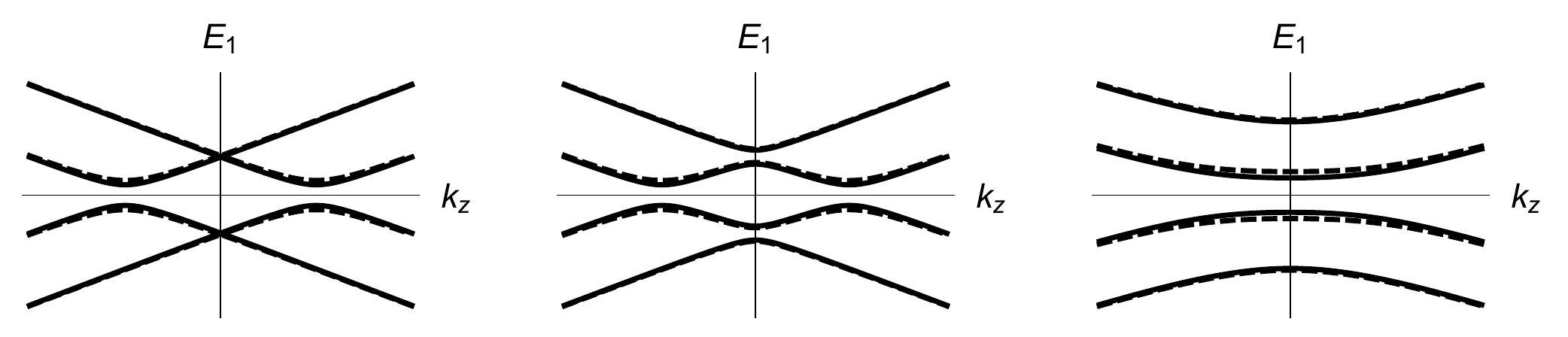}
	\caption{Energy dispersion for the first excited Landau level, $E_1$, for both spin up (black curve) and down (dashed black curve) in three different scenarios: (i) $M=0$; (ii) $\frac{M}{b}<1$ and (iii) $\frac{M}{b}>1$. Note that for the excited Landau levels ($n\geq1$) the Weyl semimetal state is destroyed. Also, there is an energy gap $\Delta E_{n}^2$ between the up and down spins, which is of order $\Delta E_{n}^2 = \beta = 2\,eB$.}
	\label{fig:ll1}
\end{figure}

Finally, in Fig.\ref{fig:ll1} we display the energy dispersion corresponding to the first excited Landau level for both spins up and down.

\begin{figure}[H]
	\centering
	\includegraphics[scale=0.7]{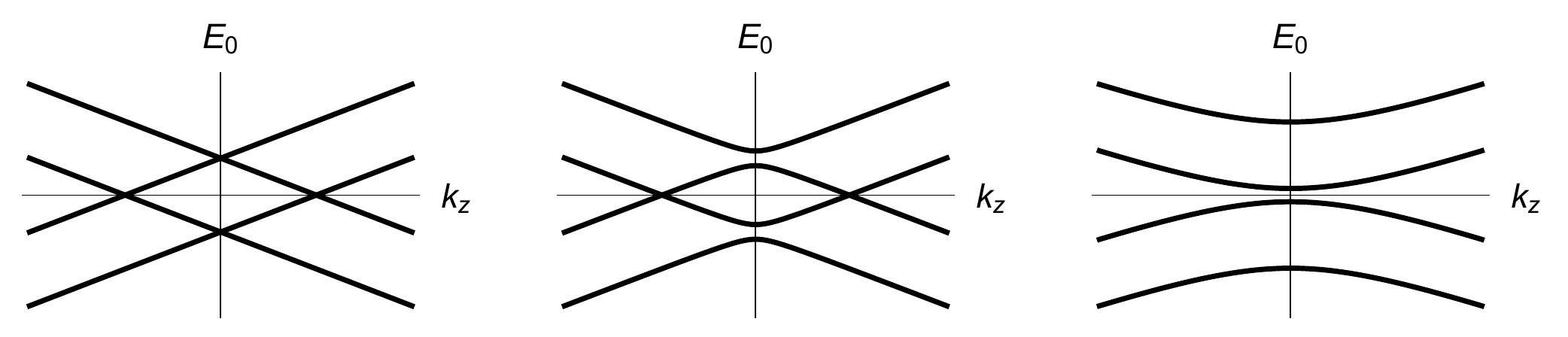}
	\caption{Energy dispersion for zero magnetic field ($B=0$) in three different scenarios: (left panel) $M=0$; (middle panel) $\frac{M}{b}<1$ and (right panel) $\frac{M}{b}>1$. Note that there is a quantum phase transition between $\frac{M}{b}<1$ and $\frac{M}{b}>1$. For $\frac{M}{b}>1$, the system is gapped and become a trivial semimetal, i.e., an insulator.}
	\label{fig:llB0}
\end{figure}

Figs.\ref{fig:ll} and \ref{fig:ll1} should be compared with the energy dispersion when the magnetic field is absent. In this case, the energy dispersion is given by
\begin{equation}
	E\big|_{B=0} = \pm\left(b\pm\sqrt{k_z^2+M^2}\right).
\end{equation}
This energy dispersion is displayed in Fig.\ref{fig:llB0}
One can see from figure \ref{fig:ll} that, even though the presence of an external magnetic field ``gaps out'' the  band touching points present in the $ B = 0 $ case (Fig.\ref{fig:llB0}), the LLL still supports a Weyl semimetal phase, once that it still hosts a linear dispersive branch with a well defined chirality close to Fermi level (set to zero for convenience) which is in fact the most general condition for the establishment of the phase, since it automatically yields a non zero flux of the corresponding Berry Curvature \cite{landsteiner_notes_2016, gao_topological_2019}.
The maintenance of Weyl points under weak, intermediate and strong magnetic fields is a far general phenomena with roots on the topological nature of Weyl points, \textit{i.e.} the fact that they are robust against perturbations that preserves the original symmetries of the system. To be more precise, in the absence of an external magnetic field perpendicular to the separation of Weyl points, the momenta along the latter it is still a good quantum number and therefore, the many body wavefunction of the electron occupying the state of a Weyl node has a definitive momenta eigenvalue \cite{lu_high-field_2015}.
The situation changes dramatically if one includes a magnetic field perpendicular to the Weyl nodes: Now the translation symmetry along the Weyl points is broken, the corresponding momenta is not a good quantum number anymore and the many body wavefunction is broadened into a finite peak around each node, allowing for the corresponding collective counter-propagating modes to hybridize and establish an electronic gap\footnote{One can borrow a classical analogue: When one introduces the magnetic field, each electron now describes a cyclotron motion in the plane perpendicular to the applied magnetic field. This motion delocalizes the electrons and their corresponding wave functions, allowing for their superposition and a consequent energy difference, i.e, an electronic gap.}. If the Weyl nodes separation is smaller and the inverse of the magnetic length $k_{ mag } =\sqrt{eB}$ the aforementioned electronic gap is not negligible, which effectively ``gaps out'' the systems leading to the magnetic tunneling effect \cite{zhang_magnetic-tunnelling-induced_2017}.

\section{Holographic Semimetal Model in a Uniform Magnetic Field}
\label{sec3}

\subsection{Holographic Setup}

The holographic model for the Weyl semimetal we consider here is an extension based on \cite{Landsteiner:2015lsa,Landsteiner:2015pdh} (for a review see, e.g, \cite{Landsteiner:2019kxb}), in which we include a background magnetic field. The bulk action is given by
\begin{equation} \label{action}
	S = \int d^{5}x\,\sqrt{-g}\Bigg[\frac{1}{2\kappa^2}\left(R + \frac{12}{L^2}\right) - \frac{1}{4}\,\left(F^2 + \mathcal{F}^2\right)
	+ \frac{\alpha}{3}\,\epsilon^{abcde}A_{a}\left(\mathcal{F}_{bc}\,\mathcal{F}_{de} + 3\,F_{bc}\,F_{de}\right) - (D\Phi)^2 - V(\Phi)\Bigg],
\end{equation}
where, $\kappa^2$ is the Newton's constant, $L$ is the AdS radius, $F = dV$ and $\mathcal{F} = dA$ are the vector and axial gauge field strengths, $\alpha$ is the Chern-Simons coupling constant, $D_{a}\Phi = (\nabla_{a}-i\,q\,A_{a})\,\Phi$ is the covariant derivative, and $\Phi$ is a complex scalar field charged under the $U(1)$ axial gauge symmetry. The scalar potential $ V(\Phi) $ is given by
\begin{equation}
	V(|\Phi|) = m^{2}\,|\Phi|^2 + \frac{\lambda}{2}\,|\Phi|^4,
\end{equation}
where, according to the holographic dictionary, $ (m\,L)^2 = \Delta(\Delta-4) $, where $\Delta$ is the scaling dimension of the operator dual to $\Phi$ and is chosen such that it has scaling dimension $\Delta=3$, which gives $m^2=-3$. Throughout this work we will work in units in which $2\kappa^2 = L = 1 $. Furthermore, we will set $\alpha=q=\lambda=1$ so that the quantum critical point at zero magnetic field will be at $\mathcal{M}_{\textrm{crit}}:= \frac{M}{b}\Big|_{{\textrm{crit}}} \approx 0.995$, where $M$ and $b$ are the mass and time-reversal breaking parameters appearing in \eqref{diraclagrangian} and, in the holographic model described by \eqref{action}, they are implemented via the following boundary conditions on the scalar and axial vector fields $\Phi$ and $A_z$:
\begin{equation}
\lim_{r\to\infty}r\Phi(r) = M; \quad \lim_{r\to\infty}A_z(r) = b.	
\end{equation} 

The field equations coming from the action \eqref{action} are given by
\begin{eqnarray}
	R_{ab} - \frac{1}{2}\,g_{ab}\,(R + 12) - T_{ab} &=& 0, \label{einseq} 
	\\
	\nabla_{b}(F^{ba}) + 2\,\alpha\,\epsilon^{abcde}\,\mathcal{F}_{bc}F_{de} &=& 0,\label{veq}
	\\
	\nabla_{b}(\mathcal{F}^{ba}) + \alpha\,\epsilon^{abcde}\,(\mathcal{F}_{bc}\mathcal{F}_{de} + F_{bc}F_{de}) -i\,q\,\left(\Phi^{*}D^{a}\Phi - (D^{a}\Phi)^{*}\Phi \right) &=& 0, \label{axeq}
	\\
	D_{a}D^{a}\Phi - \partial_{\Phi^{*}}\,V(\Phi) &=& 0 \label{scalareq}.
\end{eqnarray}
where $T_{ab}$ is the total energy-momentum tensor, which is given by
	\begin{align}
		T_{ab} =\; &\frac{1}{2}\,\Big(\mathcal{F}_{ac}\mathcal{F}_{b}^{c} - \frac{1}{4}g_{ab}\mathcal{F}^2\Big) + \frac{1}{2}\,\Big(F_{ac}F_{b}^{\; c}-\frac{1}{4}g_{ab}\,F^2\Big)\nonumber \\
		&+ \frac{1}{2}\,\Big((D_{a}\Phi)^{*}\,D_{b}\Phi + (D_{b}\Phi)^{*}\,D_{a}\Phi\Big)-\frac{1}{2}g_{ab}\Big((D_{c}\Phi)^{*}D^{c}\Phi + V(\Phi)\Big).
	\end{align}

The finite-temperature ansatz we consider for the background metric, scalar field, vector and axial vector fields are given by
\begin{eqnarray}\label{ansatz1}
	ds^2 &=& - f(r)\,dt^2 + \dfrac{dr^2}{f(r)} + u(r)\,\left(dx^2 + dy^2\right) + h(r)\,dz^2 , \nonumber\\
	\Phi &=& \Phi(r),\\ \nonumber
	V &=&  V_{t}(r)\,dt + \frac{B}{2}\,\left(- y\,dx + x\,dy\right); \quad A = 
	A_{z}(r)\,dz, 
\end{eqnarray}
where $B$ is a constant magnetic field along the $z$ direction. Using this ansatz the equations of motion for the background functions $u,f,h,A_z,V_t,\Phi$ are given by:  
\begin{eqnarray}
	\begin{aligned}\label{equ}
		\frac{u''}{u} &+ \frac{f''}{2\,f} + \left(\frac{f'}{f} - \frac{u'}{2\,u}\right)\,\frac{u'}{2\,u} - \frac{1}{4}\,\left(\frac{V_t'^2}{f} + \frac{A_z'^2}{h}\right) + \frac{1}{2}\,\Phi'^2 
		\\
		&+ \left(h\,(\lambda\,\Phi^2-6)-2\,q^2\,A_z^2\right)\,\frac{\Phi^2}{4\,h\,f} - \frac{6}{f} + \frac{B^2}{4\,u^2\,f} = 0,
	\end{aligned} 
	\\
	\frac{f''}{f} - \frac{u''}{u} + \left(\frac{f'}{f} - \frac{u'}{u}\right)\,\frac{h'}{2\,h} - \frac{V_t'^2}{f} - \frac{B^2}{f\,u^2} &=& 0 \label{eqf},
	\\
	\Phi'' + \left(\frac{f'}{f}+\frac{h'}{2\,h}+\frac{u'}{u}\right)\,\Phi' +  \left(3 - q^2\,\frac{A_z^2}{h} - \lambda\,\Phi^2\right)\,\frac{\Phi}{f} &=& 0, \label{eqphi}
	\\
	A_z'' + \left(\frac{f'}{f}-\frac{h'}{2\,h}+\frac{u'}{u}\right)\,A_z' -\frac{2\,q^2\,\Phi^2}{f}\,A_z + \frac{8\,\alpha\,B\,\sqrt{h}}{f\,u}\,V_t' &=& 0, \label{azeq}
	\\
	\left(\sqrt{h}\,u \, V_{t}' +8\,\alpha\,B A_{z}\right)'&=& 0, \label{vteq}
\end{eqnarray}
and a first-order constraint given by
\begin{eqnarray}
	\begin{aligned}
		&\left(\frac{h'}{h} + \frac{u'}{2\,u}\right)\frac{u'}{2\,u} + \left(\frac{h'}{2\,h}+\frac{u'}{u}\right)\frac{f'}{2\,f} + \frac{1}{4}
		\left(\frac{V_t'^2}{f} - \frac{A_z'^2}{h}\right) - \frac{1}{2}\,\Phi'^2
		\\
		& + \left(h\,(\lambda\,\Phi^2-6)+2\,q^2\,A_z^2\right)\,\frac{\Phi^2}{4\,h\,f} - \frac{6}{f} + \frac{B^2}{4\,u^2\,f} = 0. \label{constraint1}
	\end{aligned}
\end{eqnarray}
The prime $'$ denotes derivative with respect to the radial coordinate $r$. This system of differential equations can only be solved numerically with proper boundary conditions (see Appendix \ref{apendiceA} for more details on the asymptotic expansions and the numerical integration method). Note that an analytic solution for $V_t$ in \eqref{vteq} can be formally obtained as
\begin{align}\label{Vtsol}
	V_{t}(r) = \mu\left(1 - \dfrac{\int_{\infty}^{r}\frac{1}{u(x)\sqrt{h(x)}}\,dx}{\int_{\infty}^{r_h}\frac{1}{u(x)\sqrt{h(x)}}\,dx}\right) + 8\,\alpha\,B\Bigg(&\dfrac{\int_{\infty}^{r_h}\frac{A_{z}(x)}{\sqrt{h(x)}\,u(x)}\,dx}{\int_{\infty}^{r_h}\frac{1}{\sqrt{h(x)}\,u(x)}\,dx}\,\int_{\infty}^{r}\frac{1}{\sqrt{h(x)}\,u(x)}\,dx \nonumber\\
	&- \int_{\infty}^{r}\frac{A_{z}(x)}{\sqrt{h(x)}\,u(x)}\,dx\Bigg),
\end{align} 
where we used the boundary conditions $V_{t}(\infty)=\mu$ and $ V_{t}(r_h)=0 $ where $\mu$ is the chemical potential and $ r_h $ is the horizon position, such that $f(r_h)=0$. Since the Weyl semimetal phase is a zero density state, in this work we will take $\mu=0$. Then, the formal solution for $V_t$ becomes
\begin{equation}\label{Vtsol2}
V_{t}(r) = 8\,\alpha\,B\Bigg(\dfrac{\int_{\infty}^{r_h}\frac{A_{z}(x)}{\sqrt{h(x)}\,u(x)}\,dx}{\int_{\infty}^{r_h}\frac{1}{\sqrt{h(x)}\,u(x)}\,dx}\,\int_{\infty}^{r}\frac{1}{\sqrt{h(x)}\,u(x)}\,dx \nonumber\\
- \int_{\infty}^{r}\frac{A_{z}(x)}{\sqrt{h(x)}\,u(x)}\,dx\Bigg).
\end{equation}

\clearpage

\subsection{Numerical Results}

In this subsection we present our numerical results for the bulk profile of the background fields and for the quantum critical point (QCP) as a function of the magnetic field. The results were obtained by numerically solving the system of equations \eqref{equ}-\eqref{vteq} at small but finite temperature. The reason behind this is that the boundary conditions for the full zero-temperature solution with magnetic field are very tricky, and we will explore this analysis further in an upcoming publication.

Due to the scaling symmetries of the background, the only relevant parameters are the dimensionless ratios which we are going to define, for convenience, as follows:
\begin{eqnarray}
\mathcal{T}&:=&\frac{T}{b} = 0.05; \quad \emph{Temperature (fixed)}, \nonumber\\
\mathcal{M}&:=&\frac{M}{b}; \quad \emph{Effective coupling}, \nonumber\\
\mathcal{B}&:=&\frac{B}{T^2}; \quad \emph{Magnetic field}, \nonumber\\
\mathcal{M}_{\textrm{crit}}&:=&\left( \frac{M}{b}\right)_{\textrm{crit}} ; \quad \emph{Quantum critical point (QCP)}, \nonumber
\end{eqnarray}
with $\mathcal{M}$ and $\mathcal{B}$ being the free parameters of the model.

\subsubsection{Background Profiles}

\begin{figure}[H]
	\centering
	\includegraphics[width=0.4\linewidth]{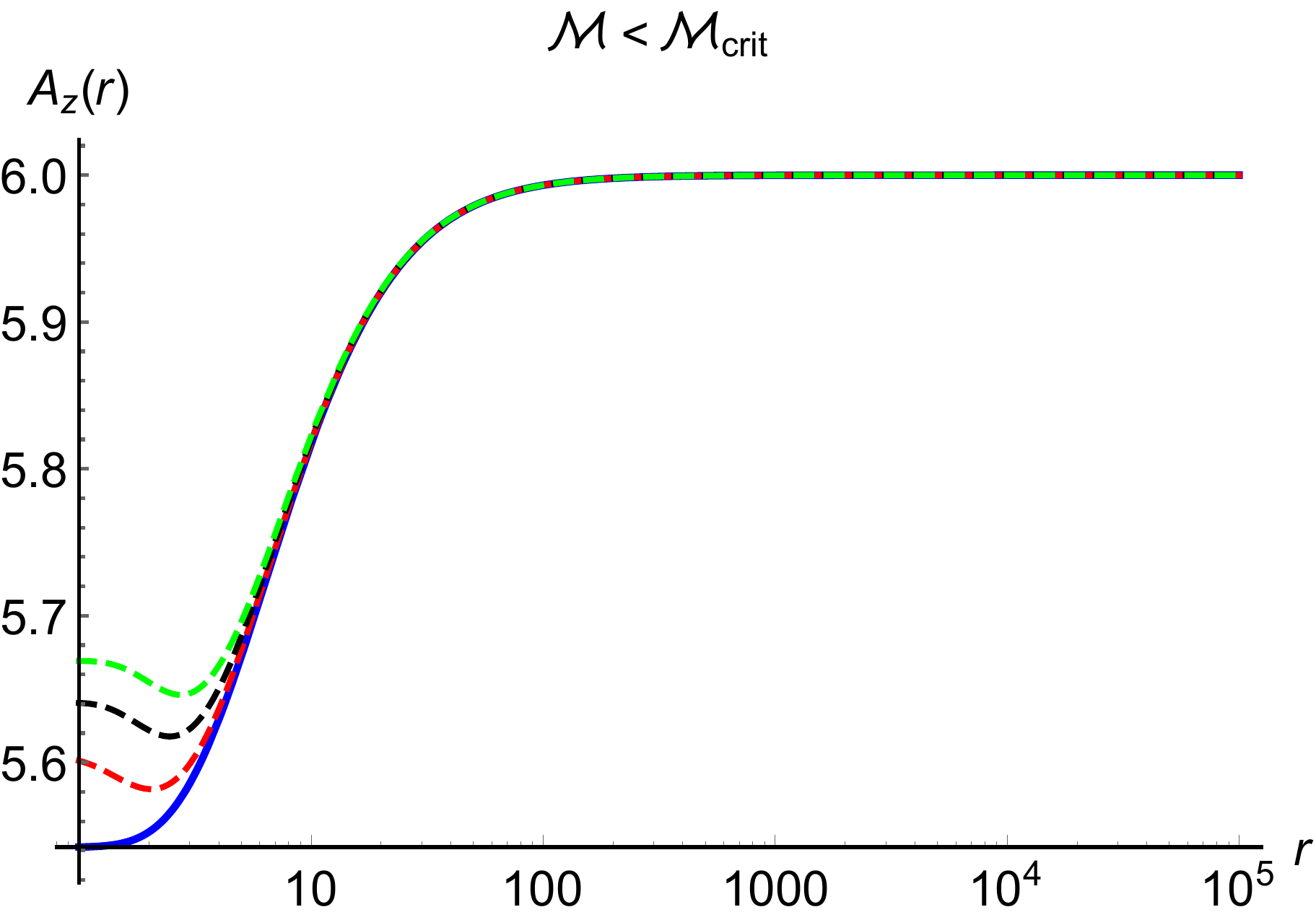}
	\includegraphics[width=0.4\linewidth]{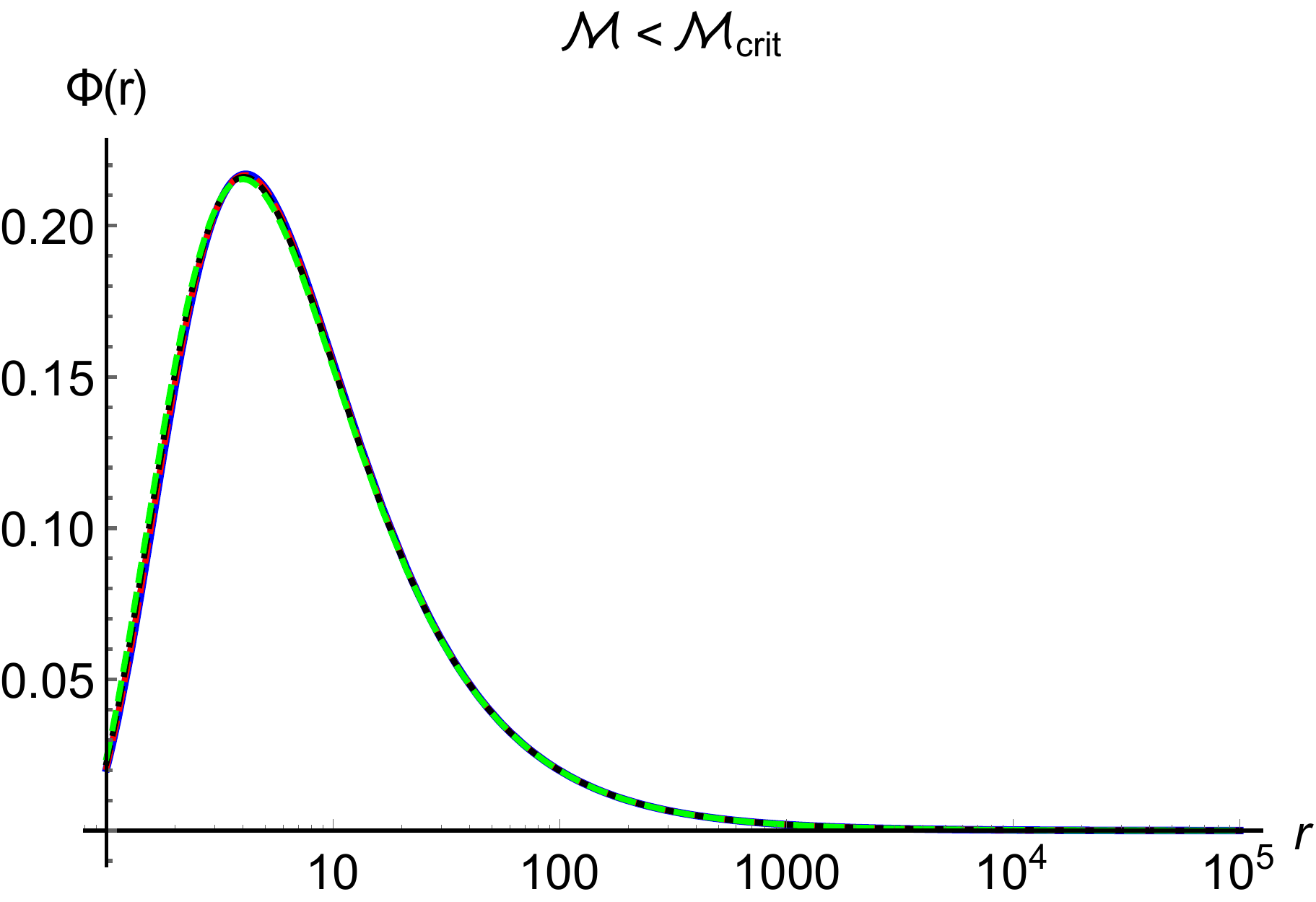}
	\caption{Bulk profile of the axial vector and scalar field in the topologically non-trivial (semimetal) phase at $\mathcal{T} = 0.05$ for different values of magnetic field: $ \mathcal{B}=0 $ (blue curve), $ \mathcal{B}=\pi^2 $ (dashed red), $ \mathcal{B}=2\,\pi^2 $ (dashed black), $ \mathcal{B}=3\,\pi^2 $ (dashed green).}
	\label{fig:az1}
\end{figure}

\begin{figure}[H]
	\centering
	\includegraphics[width=0.4\linewidth]{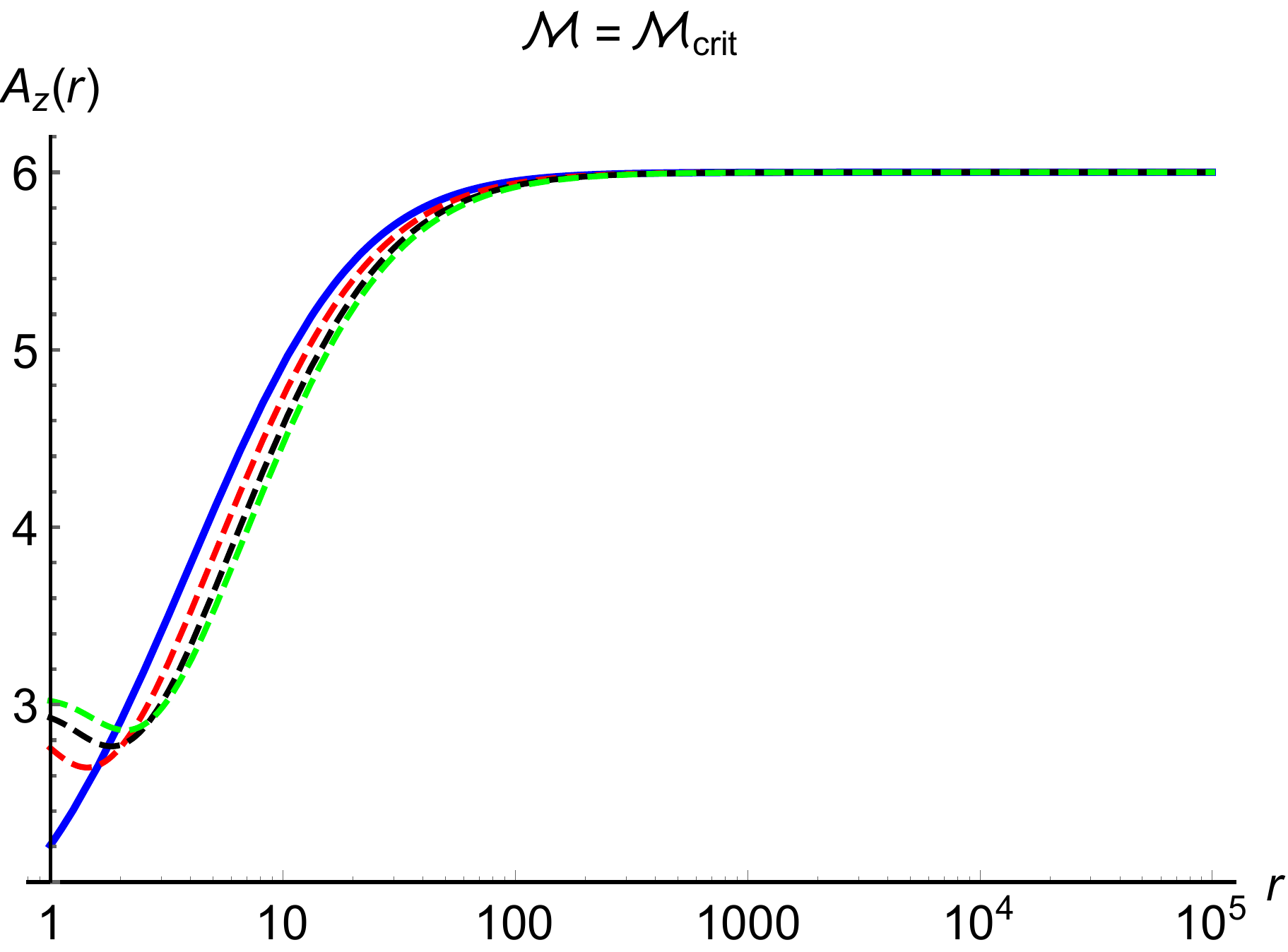}
	\includegraphics[width=0.4\linewidth]{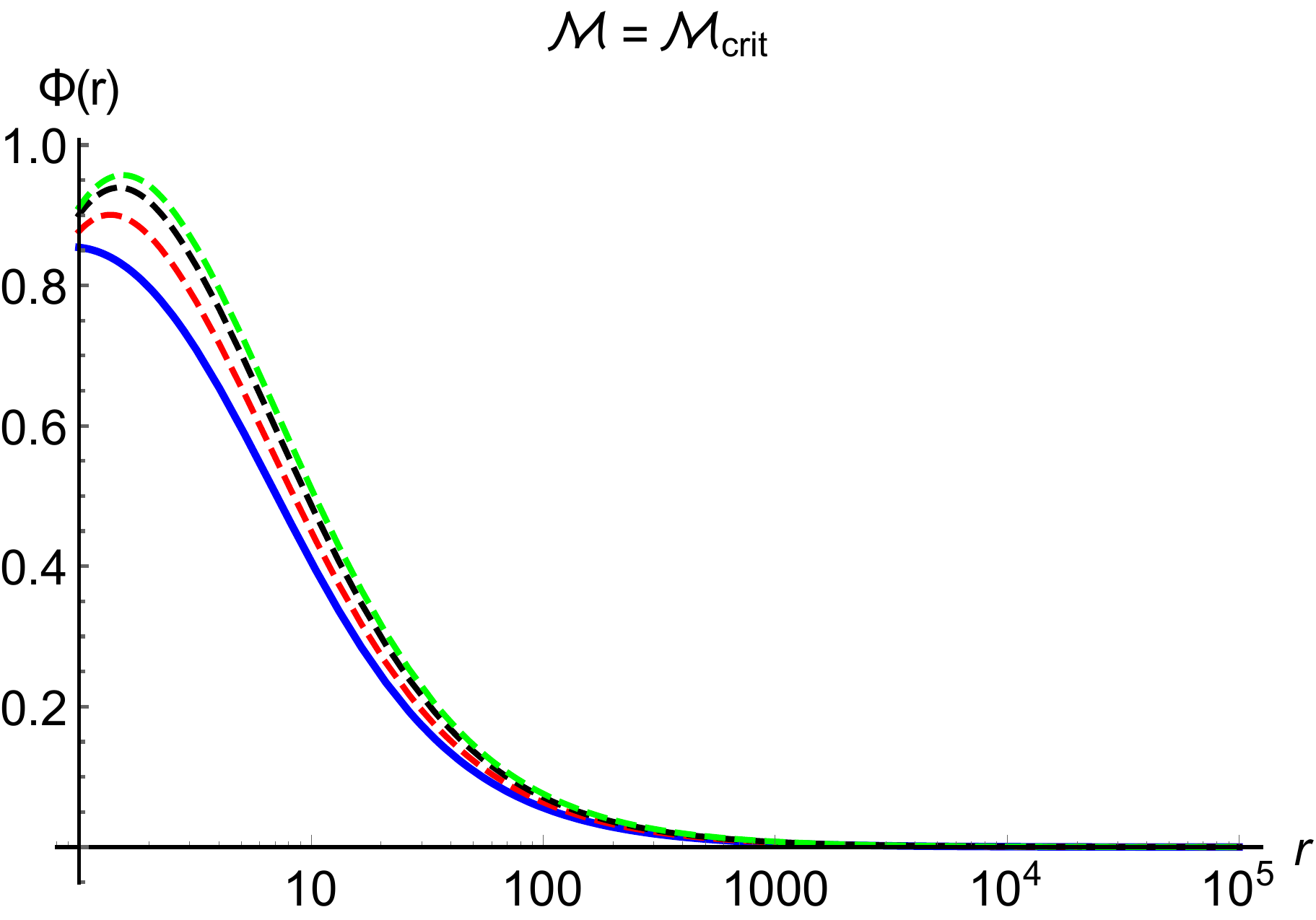}
	\caption{Bulk profile of the axial vector and scalar field in the quantum critical phase at $\mathcal{T} = 0.05$ and for different values of magnetic field: $ \mathcal{B}=0 $ (blue curve), $ \mathcal{B}=\pi^2 $ (dashed red), $ \mathcal{B}=2\,\pi^2 $ (dashed black), $ \mathcal{B}=3\,\pi^2 $ (dashed green).}
	\label{fig:az2}
\end{figure}

\begin{figure}[H]
	\centering
	\includegraphics[width=0.4\linewidth]{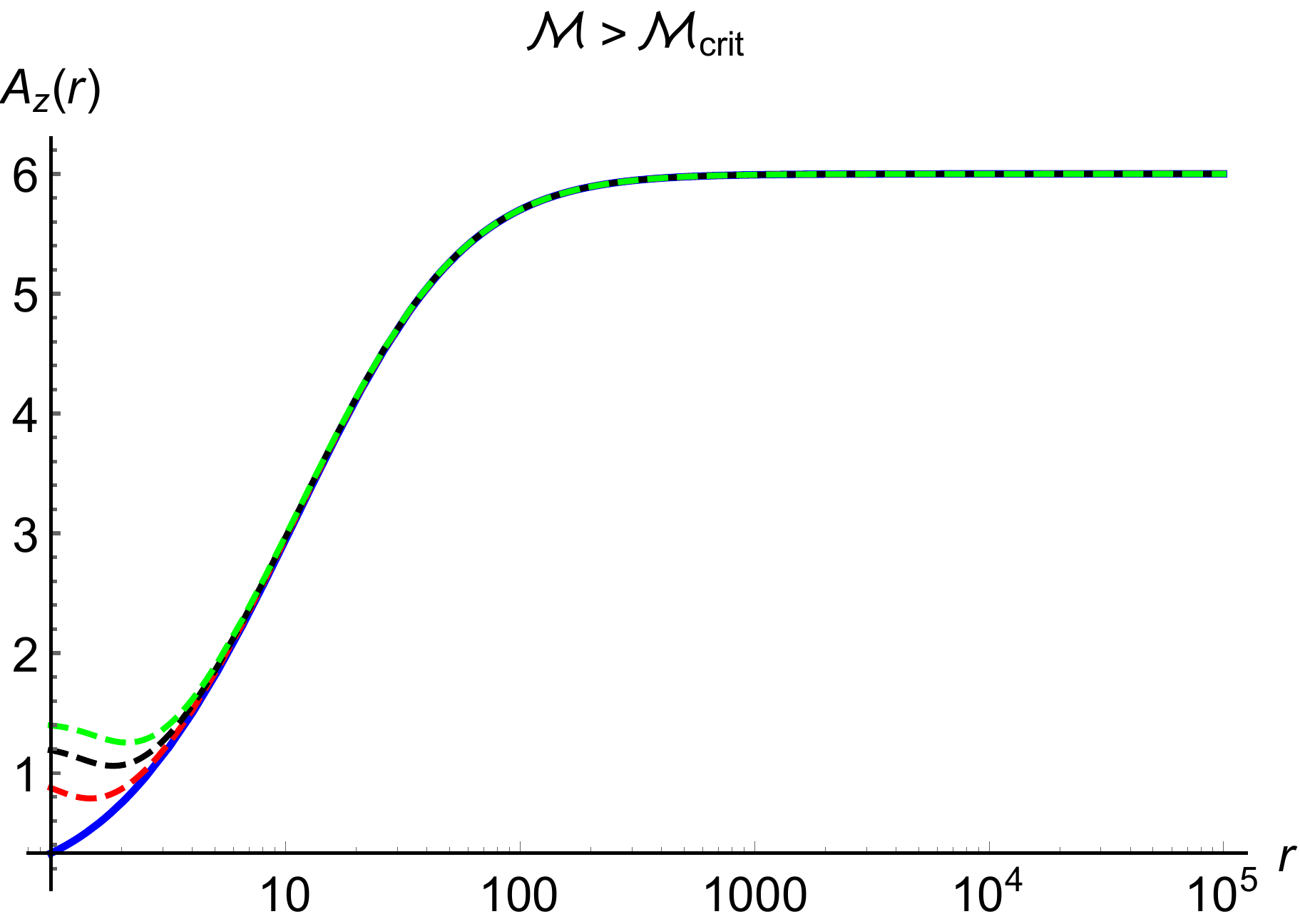}
	\includegraphics[width=0.4\linewidth]{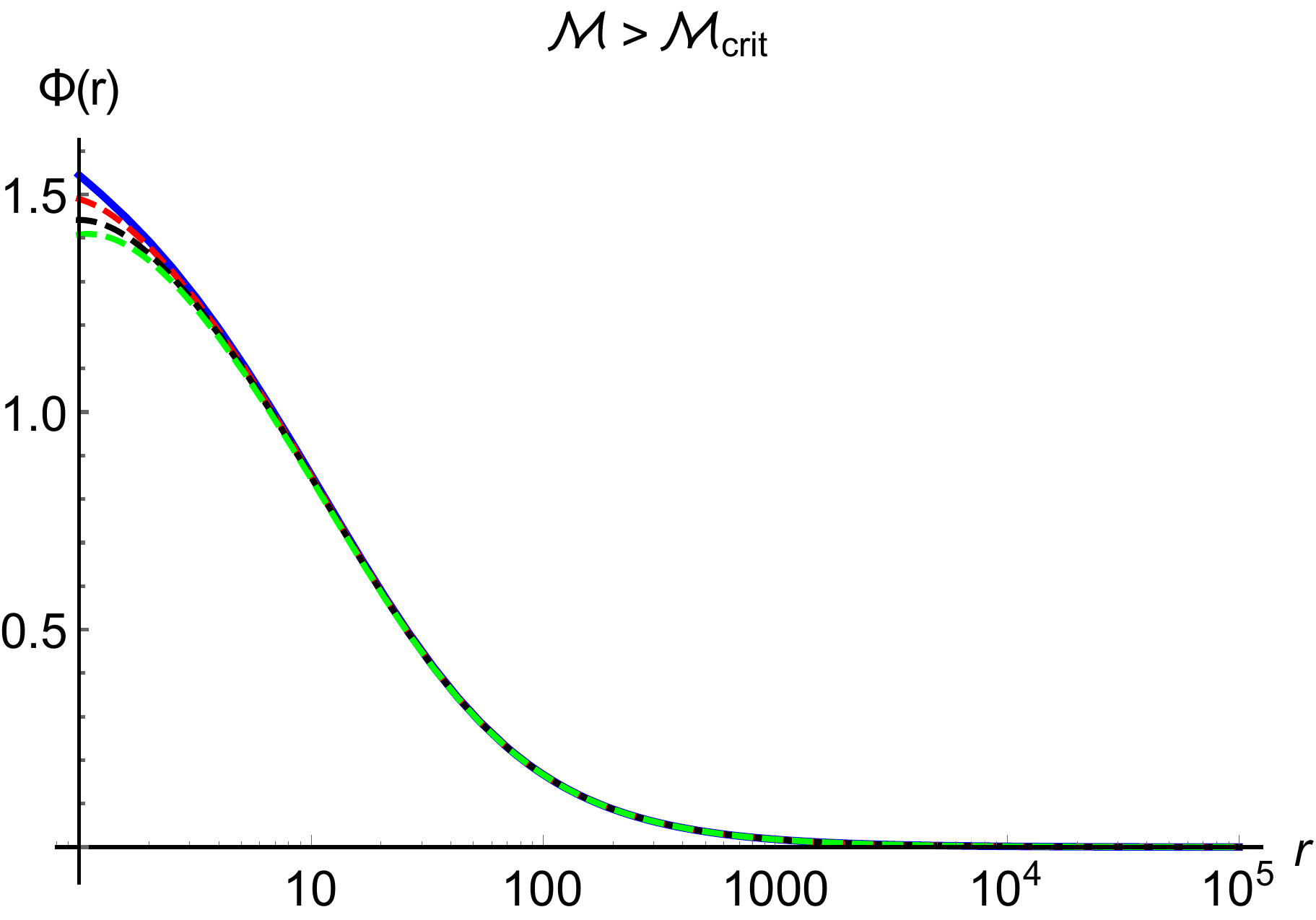}
	\caption{Bulk profile of the axial vector and scalar field in the topologically trivial (insulator) phase at $\mathcal{T} = 0.05$ and for different values of magnetic field: $ \mathcal{B}=0 $ (blue curve), $ \mathcal{B}=\pi^2 $ (dashed red), $ \mathcal{B}=2\,\pi^2 $ (dashed black), $ \mathcal{B}=3\,\pi^2 $ (dashed green).}
	\label{fig:az3}
\end{figure}

In Figs. \ref{fig:az1}, \ref{fig:az2} and \ref{fig:az3}, we display the holographic RG flow of the axial vector and scalar fields at all topological phases, namely the semimetal phase ($\mathcal{M}<\mathcal{M}_{\textrm{crit}}$), the critical phase ($\mathcal{M}=\mathcal{M}_{\textrm{crit}}$) and the topologically trivial phase ($\mathcal{M}>\mathcal{M}_{\textrm{crit}}$) for different values of $\mathcal{B}$ and small $\mathcal{M}$. Notice that the value of scalar field is almost zero at the horizon in the semimetal phase and it jumps at the critical point, and in the trivial phase it remains increasing near the horizon. Meanwhile, the axial field has a continuous behavior in all phases for all values of the magnetic field $\mathcal{B}$. However, at nonzero magnetic field, one can observe that the axial vector field develops a dip near the horizon, which makes its behavior non-monotonic for non-zero field.

\begin{figure}[H]
	\centering
	\includegraphics[width=0.4\linewidth]{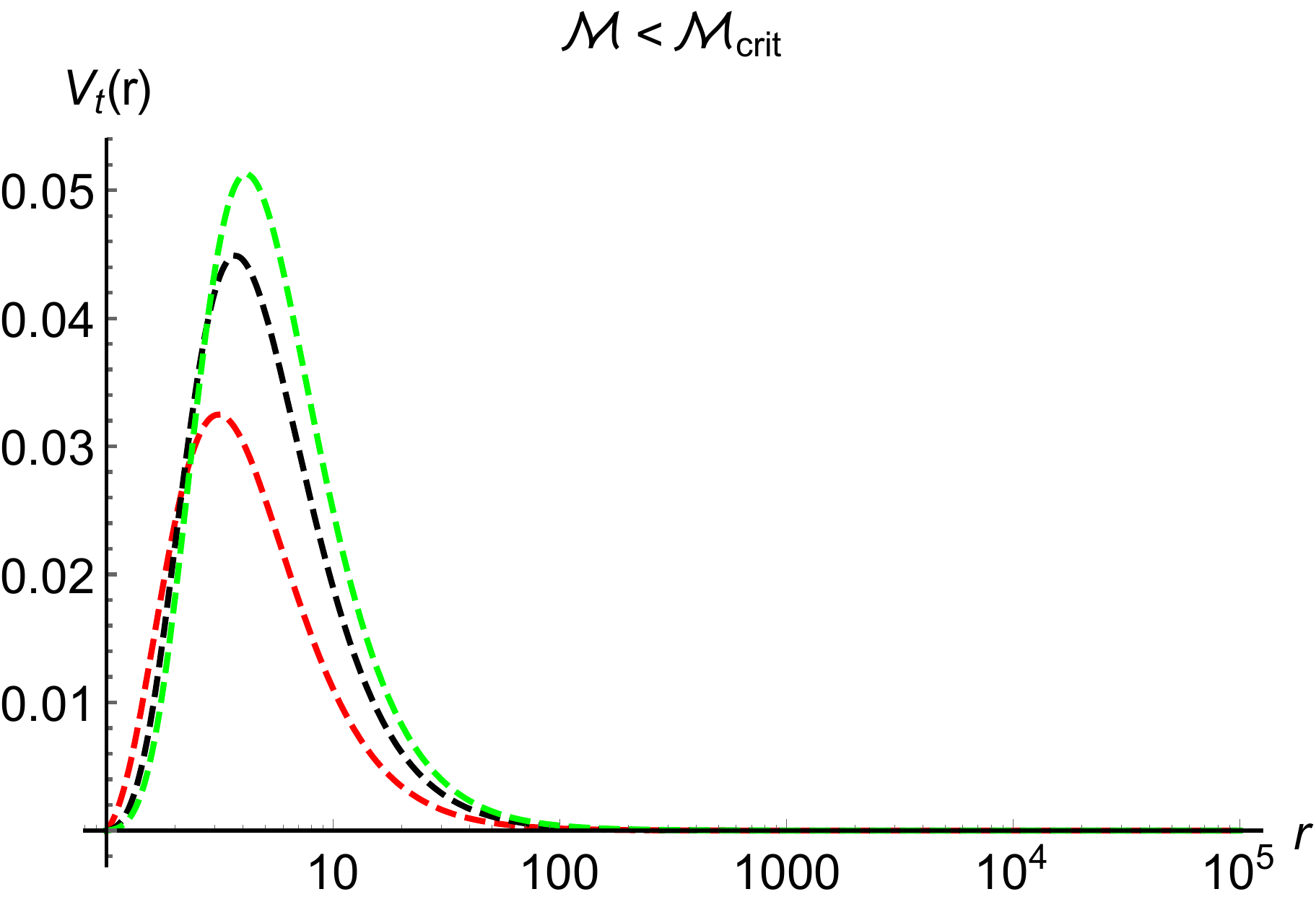}
	\includegraphics[width=0.4\linewidth]{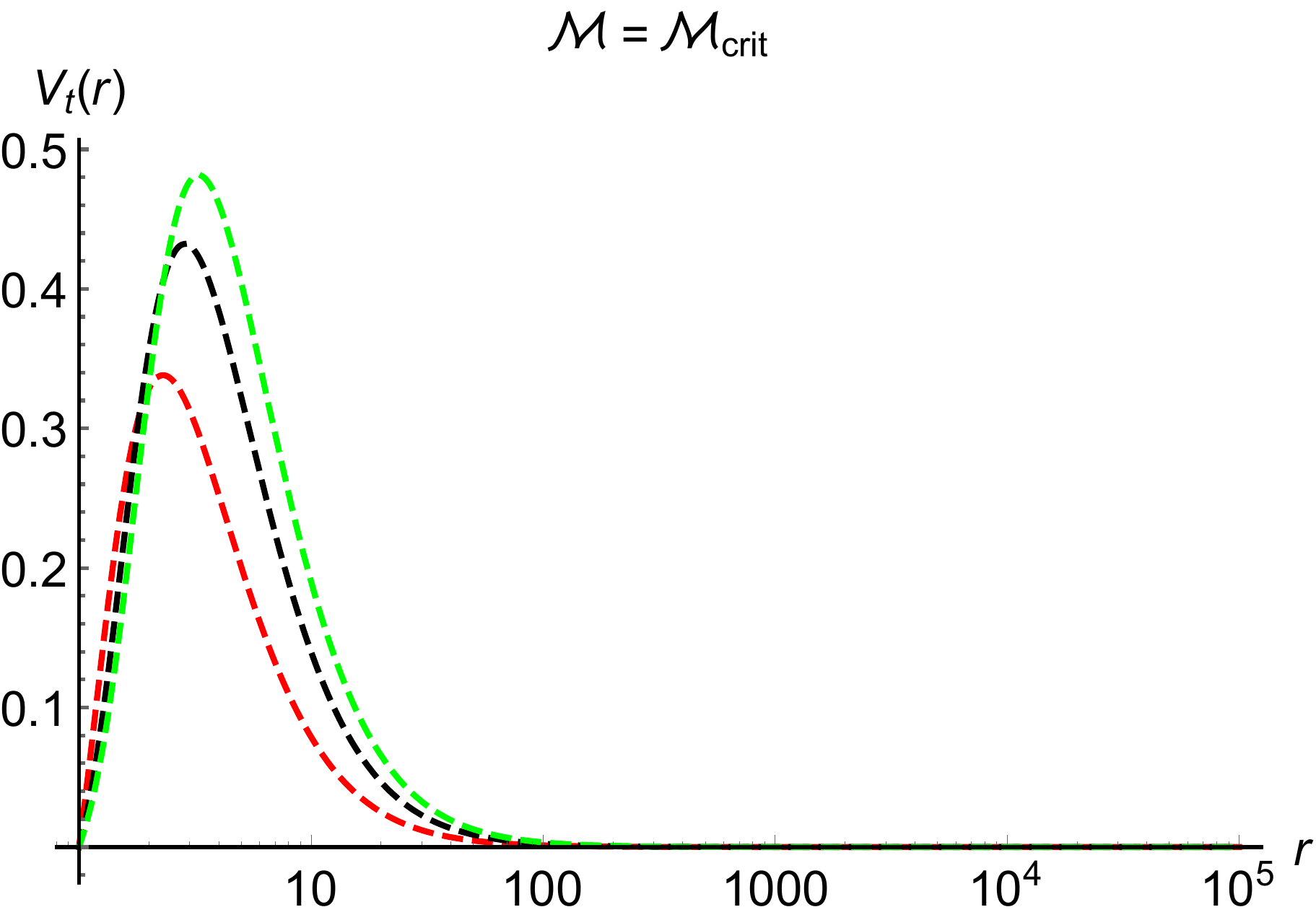}
	
	\vspace{5mm}
	
	\includegraphics[width=0.4\linewidth]{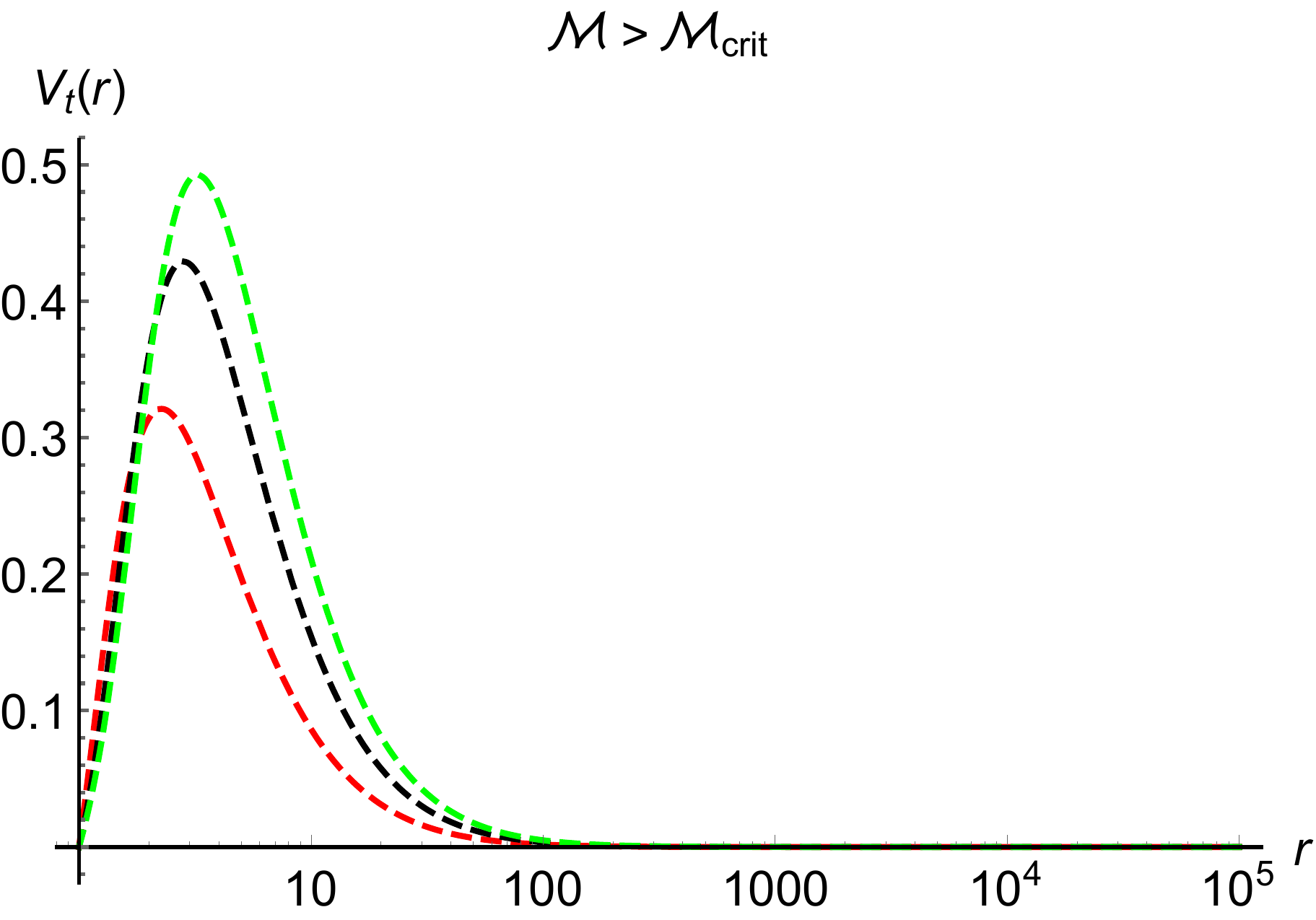}
	\caption{Bulk profile of the vector potential in all topological phases at $\mathcal{T} = 0.05$ and for different values of magnetic field: $ \mathcal{B}=\pi^2 $ (dashed red), $ \mathcal{B}=2\,\pi^2 $ (dashed black), $ \mathcal{B}=3\,\pi^2 $ (dashed green).}
	\label{fig:Vt}
\end{figure}

Finally, in Fig. \ref{fig:Vt} we display the bulk profile of the vector field in all topological phases. One can observe that it increases as we increase the magnetic field, as can be seen from the formal solution \eqref{Vtsol2} that is proportional to $B$. Also, note that the vector field has its peak increased substantially in the critical region $\mathcal{M}=\mathcal{M}_{\textrm{crit}}$, and it remains stable in the trivial phase $\mathcal{M}>\mathcal{M}_{\textrm{crit}}$. Finally, it is worthy mention that at the UV boundary the vector field is always zero because we have set the chemical potential to zero ($ \mu=0 $), i.e, there is no electric charge density.  

\subsubsection{QCP as a function of $B$}

It has been known that the order parameter of the quantum phase transition is the anomalous Hall conductivity $\sigma_{AHE}$, which was obtained in holography in terms of the horizon data \cite{Landsteiner:2015lsa,Landsteiner:2015pdh}. Thus, the quantum critical point (QCP) $\mathcal{M}_{\textrm{crit}}$ can be extracted from the behavior of $\sigma_{AHE}$ as a function of the effective coupling $\mathcal{M}$ at both zero and finite temperature. However, in the presence of the magnetic field we found more convenient to use another probe of the QCP, which is the spatial anisotropy along the $ z $ direction in terms of the horizon data. It is defined as follows:
\begin{equation}
\epsilon := \frac{h(r_h)}{u(r_h)}-1.
\end{equation}
The anisotropy parameter $\epsilon$ as a function of the effective coupling $\mathcal{M}$ has proven to be a good probe to detect the location of the QCP as it is peaked around it, suggesting it is a point or region (in the case of finite temperature) of maximum anisotropy and strong divergences at zero temperature, as shown in \cite{baggioli_conjecture_2018}.

\begin{figure}[H]
	\centering
	\includegraphics[width=0.6\linewidth]{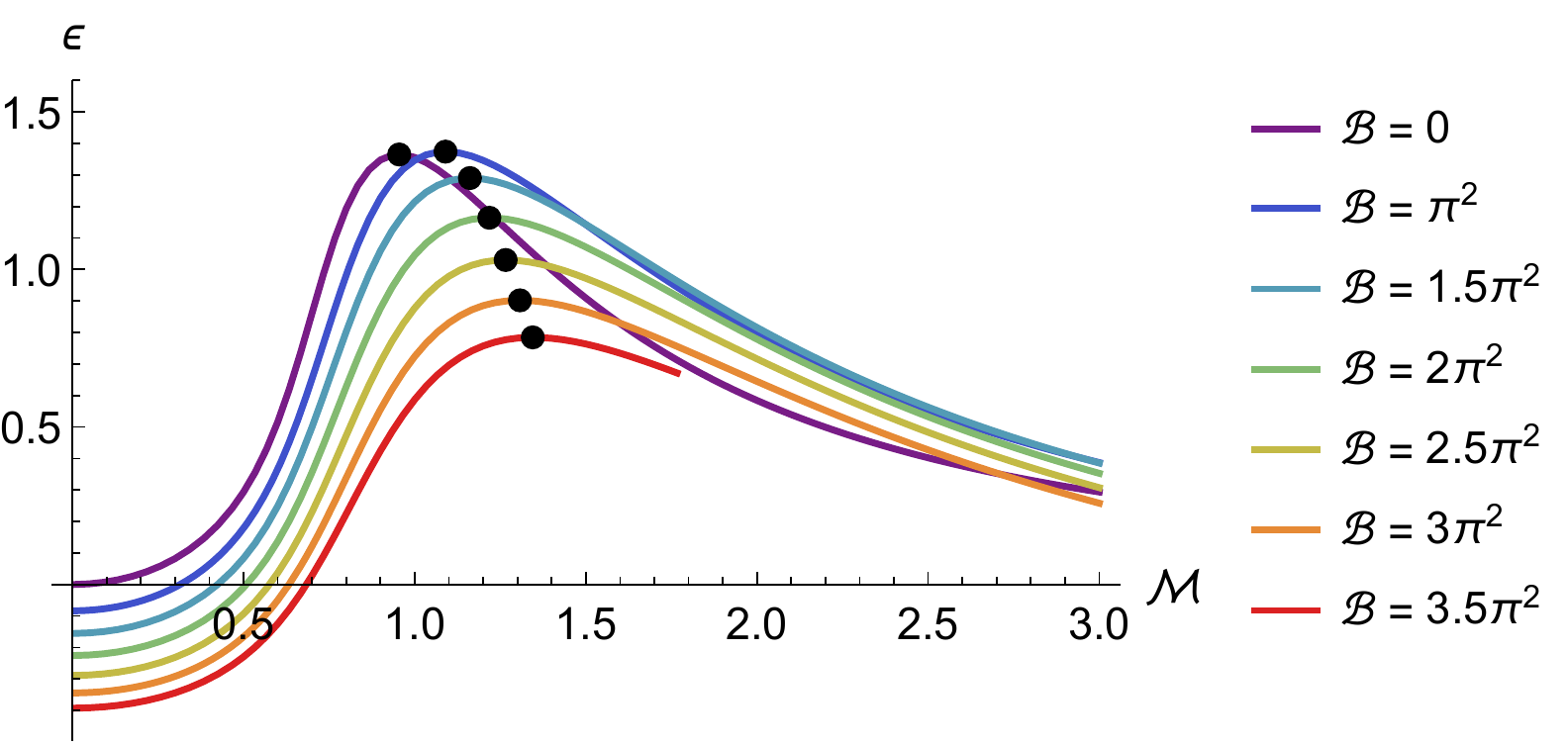}
	\caption{Horizon spatial anisotopry parameter $\epsilon$ as a function of the effective coupling $\mathcal{M}$ for different values of the magnetic field $\mathcal{B}$ at $\mathcal{T} = 0.05$. The black dots represent the QCP for each value magnetic field.}
	\label{fig:main1}
\end{figure}

From the numerical solution of the background we extracted the anisotropy parameter $\epsilon$ as a function of the effective coupling $\mathcal{M}$ for several values magnetic field $\mathcal{B}$, as displayed in Fig. \ref{fig:main1}. From thoses anisotropy curves we extracted the location of the QCP for each magnetic field by computing the value of $\mathcal{M}$ for which $\epsilon$ is maximum. These points are represented by the black dots in Fig. \ref{fig:main1}. From it one can clearly see that the QCP increases as we increase the value of $\mathcal{B}$.

\begin{figure}[H]
	\centering
	\includegraphics[width=0.4\linewidth]{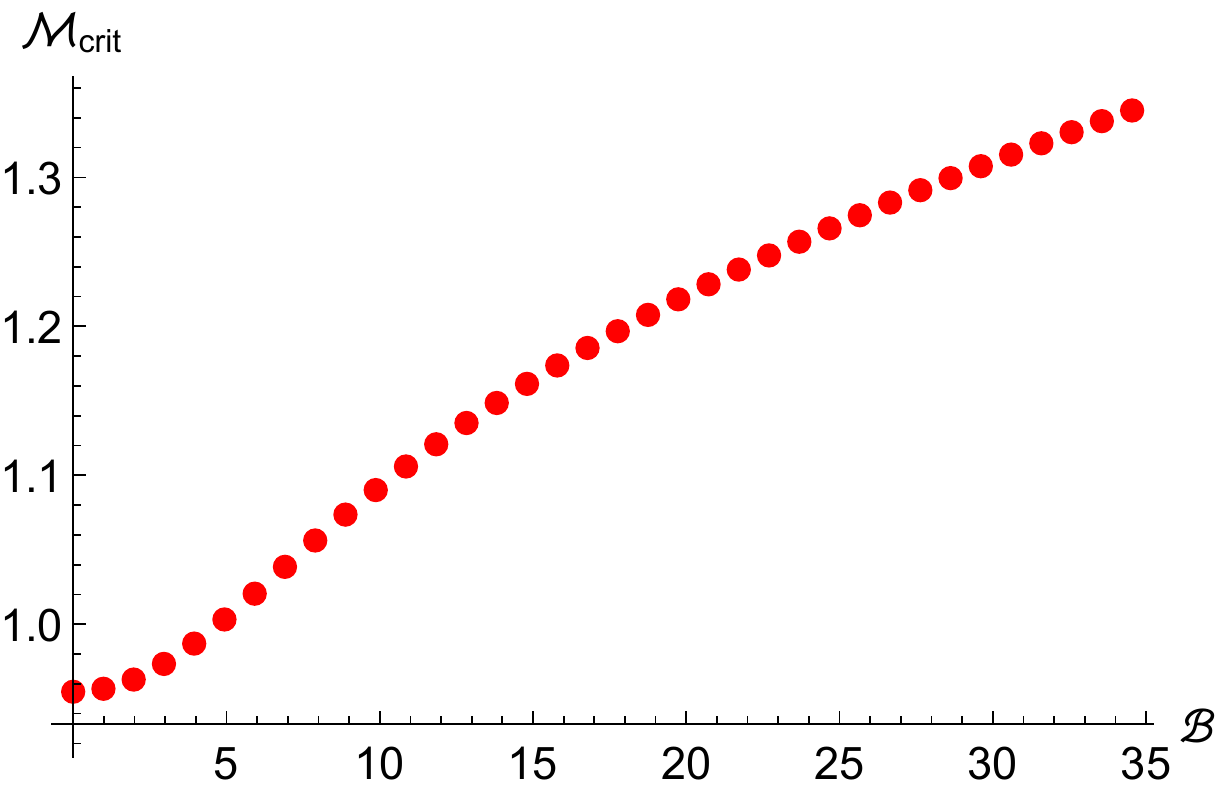}
	\includegraphics[width=0.4\linewidth]{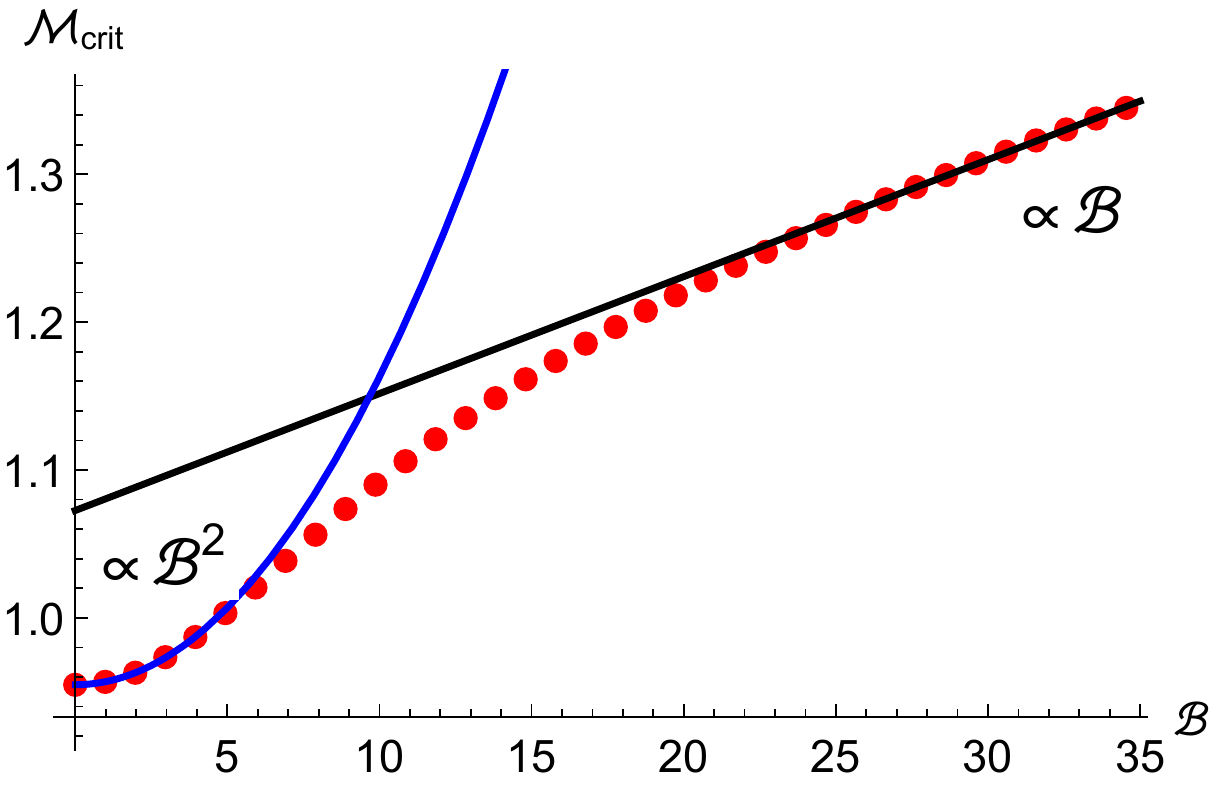}
	\caption{\textbf{Left Panel}: Quantum Critical Point (QCP), $\mathcal{M}_{\textrm{crit}}$, as a function of the magnetic field $\mathcal{B}$ at fixed small temperature $\mathcal{T} = 0.05$. \textbf{Right Panel}: Quantum Critical Point (QCP), $\mathcal{M}_{\textrm{crit}}$, as a function of the magnetic field $\mathcal{B}$ at fixed small temperature $\mathcal{T} = 0.05$ together with a quadratic-in-$ \mathcal{B} $ and linear-in-$\mathcal{B}$ fittings}
	\label{fig:main2}
\end{figure}

In order to comprehend in more detail how the QCP behaves as a function of the magnetic field, in the left panel of Fig. \ref{fig:main2} we display the behavior of $\mathcal{M}_{crit}$ as a function of $\mathcal{B}$. Note that it presents different scalings depending on the regime of the magnetic field $\mathcal{B}$. For instance, at small $\mathcal{B}$ it has a quadratic-in-$\mathcal{B}$ dependence, while for large $\mathcal{B}$ it has a linear-in-$\mathcal{B}$ dependence, as shown in the fittings presented in the right panel of Fig. \ref{fig:main2}. Those behaviors seem to agree with the semiclassical and ultraquantum magnetotransport regimes previously described in the literature within the framework of Boltzmann transport theory (see \cite{PhysRevResearch.2.033511} and references therein). It is interesting to see that this holographic model can capture those behaviors through the QCP. Ultimately, they are manifestations of quantum anomalies, notably the chiral anomaly which is responsible for the quadratic-in-$\mathcal{B}$ in the magnetoconductivity, as experimentally shown in \cite{Li_2016}.
However, in the intermediate range of $\mathcal{B}$, it should be expected a quantum oscillatory behavior periodic in $ 1/\mathcal{B}$ in the magnetotransport coefficients as described in \cite{PhysRevResearch.2.033511}, in which the present holographic model is not capable of capture it. For a more recent discussion on the role played by the quantum oscillations within the nonlinear response regime in WSMs see \cite{Zeng_2023}.

\section{Conclusion and Discussion}
\label{sec4}

In this work, we have considered an extension of the holographic Weyl semimetal model and its quantum phase transition to a trivial semimetal by including a finite magnetic field at small but finite-temperature in order to investigate the effects of the magnetic field on the quantum critical point. For this purpose we have made use of the spatial anisotropy parameter in terms of the horizon data and, from it, we extracted the QCP, $\mathcal{M}_{\textrm{crit}}$, as a function of the magnetic field $\mathcal{B}$. We have found that $\mathcal{M}_{\textrm{crit}}(\mathcal{B})$ displays a quadratic-in-$\mathcal{B}$ for weak field and linear-in-$\mathcal{B}$ for strong field compatible with results found in the literature within the Boltzmann transport theory. Since we have considered a finite but small temperature holographic model, it would be interesting  to try to construct a background at zero temperature in the presence of magnetic field \cite{Sun:2016gpy} and compare the results with the ones obtained here for $\mathcal{M}_{\textrm{crit}}(\mathcal{B})$. Furthermore, one could study the effects of the magnetic field on the QCP by analyzing other probes which have been proposed recently, such as the $ c $-function \cite{Baggioli:2020cld} and the entanglement entropy (EE) for the case of nodal line WSMs \cite{Baggioli:2023ynu}.
	
Despite the fact the electrical transport coefficients, such as the electrical conductivity, has been extensively studied in holographic semimetal model \cite{Landsteiner:2015pdh,Jacobs:2015fiv,Grignani:2016wyz,Copetti:2016ewq,Rodgers:2021azg,Zhao:2021pih}, much less attention has been given to holographic calculation of magnetotransport in WSMs. From the experimental observations, we know that the longitudinal magnetoconductivity in the Weyl semimetal is enhanced by the chiral anomaly, which is called as positive magnetoconductivity or, equivalently, negative magnetoresesivity. Experimentally, it was observed that the longitudinal conductivity in this topological material has a $B^2$ dependence at small magnetic field as a consequence of the chiral magnetic effect (CME), which is the generation of a current along the direction of the magnetic field. Therefore, it would be interesting as a possible extension of this work to consider the effects of including vector and chiral chemical potentials in the current holographic WSM model, in order to study the magnetoconductivity and see if one can observe the presence of the CME.
		
Finally, another possible extension of the this work is the study of the role played by the mixed-chiral gravitational anomaly, which yields a non-conservation of the chiral charge and energy \cite{PhysRevResearch.2.033511,PhysRevResearch.2.013088,Das:2023onx}. This anomaly leads to the enhancement of the longitudinal magneto-thermal conductivity and magneto-thermoelectric transport coefficients \cite{Chernodub:2021nff}. Thus, it would be interesting to consider this term in the holographic model, as done in \cite{Ji:2019pxx} to study the chiral-vortical conductivity in holography, and investigate the magneto-thermoelectric transport coefficients and compare with experimental predictions \cite{Gooth_2017,Schindler:2018wrd,Vu_2021}. We will leave these further studies for the future.

\medbreak

\paragraph*{\textbf{Acknowledgments}:} We would like to thank Yan Liu for useful discussions and motivation in the early stage of this project and for providing us interesting references on experimental results in topological materials. D. M. R. is supported by grant \raisebox{\depth}{\#}2021/01565-8, São Paulo Research Foundation (FAPESP).

\appendix

\section{IR and UV Asymptotic Expansions and Numerics}
\label{apendiceA}
Here we provide the asymptotic expansions for the background fields near the horizon $r\to r_h$ (IR) and near the boundary $r\to\infty$ (UV) as well as we briefly explain how we numerically integrate the equations of motion \eqref{equ}-\eqref{vteq}. 


Near the horizon we assume the background fields can be expressed in a regular Taylor expansion around $r=r_h$, which can be written as
\begin{eqnarray}
	A_z &=& A_0 + \frac{-4 \alpha B \sqrt{h_0} V_1 + A_0 q^2 u_0 \Phi_0^2}{2 \pi T  u_0}\left(r-r_h\right)+...\,,\\
	V_t &=& V_1 \left(r-r_h\right)+ ...\,,\\
	\Phi &=&\Phi_0+\frac{\Phi_0 \left(A_0^2 q^2+h_0(-3+\lambda \Phi_0^2) \right)}{4  \pi T h_0}\left(r-r_h\right)+ ...\,,\\
	f &=& 4 \pi T \left(r-r_h\right) +...\,,\\
	u &=& u_0+\frac{2 B^2 +u_0^2 \left( -24+ V_1^2 + \Phi_0^2 \left(-6 +\lambda \Phi_0^2\right) \right)}{12 \pi T u_0} \left( r-r_h\right)+...\,,\\
	h &=& h_0+\frac{B^2 h_0 -u_0^2 \left(6 A_0^2 q^2 \Phi_0^2  +h_0 \left[ -24+V_1^2+\Phi_0^2 \left(-6 +\lambda \Phi_0^2\right) \right] \right)}{12 \pi T u_0^2} \left(r-r_h\right)+...\,
\end{eqnarray}
where in the numerics we set $r_h = 1$ and $T = \frac{1}{\pi}$, and the shooting parameters near the horizon are $A_0$, $V_1$, $ \Phi_0 $, $u_0$ and $h_0$.

On the other hand, near the boundary $r\to\infty$ we impose that the background is asymptotically AdS. Thus, the asymptotic expansion for the various fields can be written as
\begin{eqnarray}
	A_z &=& b + \frac{b M^2q^2\left(9 b^2 M^2 q^2-192 \alpha ^2 B^2+M^4 \left(9 \lambda +9
		q^2+5\right)\right)}{12\left(M^4 \left(q^2+1\right)-64 \alpha ^2 B^2\right)r^2}-\frac{b
		M^2 q^2 \ln (r)}{r^2}+...\,,\\
	V_t &=& \mu + \frac{2\alpha b B M^2 q^2 \left(9 b^2 q^2+M^2 \left(9 \lambda +6 q^2+2\right)\right)}{\left(192 \alpha ^2 B^2-3 M^4
		\left(q^2+1\right)\right)\,r^2}-\frac{2 \alpha  b B M^2 q^2\ln (r)}{r^4} +...\,,\\
	\Phi &=& \dfrac{M}{r}-\frac{M \left(3 b^2 q^2+(3 \lambda +2) M^2\right)\ln (r)}{6 r^3} +...\,,\\
	f &=& r^2 -\frac{M^2}{3} + \frac{\left(-3 B^2+3 \lambda  M^4+2 M^4\right)\ln (r)}{18 r^2} +...\,,\\
	u &=& r^2 -\frac{M^2}{3} + \frac{\left(3 B^2+6 \lambda  M^4+4 M^4\right)\ln(r)}{36 r^2}+...\,,\\
	h &=& r^2 -\frac{M^2}{3} + \frac{9 b^2 M^2 q^2+(9 \lambda +14) M^4}{72
		r^2} + \frac{\left(9 b^2 M^2 q^2-3 B^2+(3 \lambda +2) M^4\right)\ln (r)}{18 r^2}.
\end{eqnarray}
Since we have considered in this work the case of zero chemical potential, we have set $\mu=0$ in numerics.
In addition, due to the underlying conformal symmetry of the background, the relevant model parameters are the dimensionless ratios previously defined in the main text $\left(T/b:= \mathcal{T}; M/b:=\mathcal{M}; B/T^2:=\mathcal{B}\right)$.

The numerical integration of the background equations of motion \eqref{equ}-\eqref{vteq} were obtained using a matching technique (for more details and practical examples we refer the reader to \cite{Andrade:2017jmt,Baggioli_2019}). In the following, we briefly summarize the numerical procedure:
\begin{itemize}
	
	\item{One solves numerically the equations from the boundary to an intermediate point by imposing the asymptotic expansion at boundary;}
	
	\item{One solves numerically the equations from the horizon to an intermediate point by imposing the asymptotic expansion at horizon;}
	
	\item{One matches the numerical solutions from the first and second steps at the intermediate point using a $\mathtt{FindRoot}$ routine in $ \mathtt{Mathematica} $. This will give five equations ($\Psi^{i}_{r\to\infty}-\Psi^{i}_{r\to r_h}$, where $\Psi^i = A, V_t, \Phi, u, h$) for the five parameters, namely $A_0$, $V_1$, $ \Phi_0 $, $u_0$ and $h_0$. For fixed $\mathcal{T}$, these parameters will vary as we vary the effective coupling $\mathcal{M}$ and the magnetic field $\mathcal{B}$.}
\end{itemize}
One could also numerically solve the equations of motions by making use of the scaling symmetries presented in the background (for more details see for instance \cite{Landsteiner:2015pdh,Liu:2020ymx})

\newpage
\bibliographystyle{apsrev4-2}
\bibliography{refs}

\begin{thebibliography}{94}%
\makeatletter
\providecommand \@ifxundefined [1]{%
 \@ifx{#1\undefined}
}%
\providecommand \@ifnum [1]{%
 \ifnum #1\expandafter \@firstoftwo
 \else \expandafter \@secondoftwo
 \fi
}%
\providecommand \@ifx [1]{%
 \ifx #1\expandafter \@firstoftwo
 \else \expandafter \@secondoftwo
 \fi
}%
\providecommand \natexlab [1]{#1}%
\providecommand \enquote  [1]{``#1''}%
\providecommand \bibnamefont  [1]{#1}%
\providecommand \bibfnamefont [1]{#1}%
\providecommand \citenamefont [1]{#1}%
\providecommand \href@noop [0]{\@secondoftwo}%
\providecommand \href [0]{\begingroup \@sanitize@url \@href}%
\providecommand \@href[1]{\@@startlink{#1}\@@href}%
\providecommand \@@href[1]{\endgroup#1\@@endlink}%
\providecommand \@sanitize@url [0]{\catcode `\\12\catcode `\$12\catcode
  `\&12\catcode `\#12\catcode `\^12\catcode `\_12\catcode `\%12\relax}%
\providecommand \@@startlink[1]{}%
\providecommand \@@endlink[0]{}%
\providecommand \url  [0]{\begingroup\@sanitize@url \@url }%
\providecommand \@url [1]{\endgroup\@href {#1}{\urlprefix }}%
\providecommand \urlprefix  [0]{URL }%
\providecommand \Eprint [0]{\href }%
\providecommand \doibase [0]{https://doi.org/}%
\providecommand \selectlanguage [0]{\@gobble}%
\providecommand \bibinfo  [0]{\@secondoftwo}%
\providecommand \bibfield  [0]{\@secondoftwo}%
\providecommand \translation [1]{[#1]}%
\providecommand \BibitemOpen [0]{}%
\providecommand \bibitemStop [0]{}%
\providecommand \bibitemNoStop [0]{.\EOS\space}%
\providecommand \EOS [0]{\spacefactor3000\relax}%
\providecommand \BibitemShut  [1]{\csname bibitem#1\endcsname}%
\let\auto@bib@innerbib\@empty
\bibitem [{\citenamefont {Schnyder}\ \emph {et~al.}(2008)\citenamefont
  {Schnyder}, \citenamefont {Ryu}, \citenamefont {Furusaki},\ and\
  \citenamefont {Ludwig}}]{schnyder_classification_2008}%
  \BibitemOpen
  \bibfield  {author} {\bibinfo {author} {\bibfnamefont {A.~P.}\ \bibnamefont
  {Schnyder}}, \bibinfo {author} {\bibfnamefont {S.}~\bibnamefont {Ryu}},
  \bibinfo {author} {\bibfnamefont {A.}~\bibnamefont {Furusaki}},\ and\
  \bibinfo {author} {\bibfnamefont {A.~W.~W.}\ \bibnamefont {Ludwig}},\ }\href
  {https://doi.org/10.1103/PhysRevB.78.195125} {\bibfield  {journal} {\bibinfo
  {journal} {Phys. Rev. B}\ }\textbf {\bibinfo {volume} {78}},\ \bibinfo
  {pages} {195125} (\bibinfo {year} {2008})}\BibitemShut {NoStop}%
\bibitem [{\citenamefont {Chiu}\ \emph {et~al.}(2016)\citenamefont {Chiu},
  \citenamefont {Teo}, \citenamefont {Schnyder},\ and\ \citenamefont
  {Ryu}}]{chiu_classification_2016}%
  \BibitemOpen
  \bibfield  {author} {\bibinfo {author} {\bibfnamefont {C.-K.}\ \bibnamefont
  {Chiu}}, \bibinfo {author} {\bibfnamefont {J.~C.}\ \bibnamefont {Teo}},
  \bibinfo {author} {\bibfnamefont {A.~P.}\ \bibnamefont {Schnyder}},\ and\
  \bibinfo {author} {\bibfnamefont {S.}~\bibnamefont {Ryu}},\ }\href
  {https://doi.org/10.1103/RevModPhys.88.035005} {\bibfield  {journal}
  {\bibinfo  {journal} {Rev. Mod. Phys.}\ }\textbf {\bibinfo {volume} {88}},\
  \bibinfo {pages} {035005} (\bibinfo {year} {2016})}\BibitemShut {NoStop}%
\bibitem [{\citenamefont {Kane}\ and\ \citenamefont
  {Mele}(2005)}]{kane_z_2_2005}%
  \BibitemOpen
  \bibfield  {author} {\bibinfo {author} {\bibfnamefont {C.~L.}\ \bibnamefont
  {Kane}}\ and\ \bibinfo {author} {\bibfnamefont {E.~J.}\ \bibnamefont
  {Mele}},\ }\href {https://doi.org/10.1103/PhysRevLett.95.146802} {\bibfield
  {journal} {\bibinfo  {journal} {Phys. Rev. Lett.}\ }\textbf {\bibinfo
  {volume} {95}},\ \bibinfo {pages} {146802} (\bibinfo {year}
  {2005})}\BibitemShut {NoStop}%
\bibitem [{\citenamefont {Fu}\ \emph {et~al.}(2007)\citenamefont {Fu},
  \citenamefont {Kane},\ and\ \citenamefont {Mele}}]{fu_topological_2007}%
  \BibitemOpen
  \bibfield  {author} {\bibinfo {author} {\bibfnamefont {L.}~\bibnamefont
  {Fu}}, \bibinfo {author} {\bibfnamefont {C.~L.}\ \bibnamefont {Kane}},\ and\
  \bibinfo {author} {\bibfnamefont {E.~J.}\ \bibnamefont {Mele}},\ }\href
  {https://doi.org/10.1103/PhysRevLett.98.106803} {\bibfield  {journal}
  {\bibinfo  {journal} {Phys. Rev. Lett.}\ }\textbf {\bibinfo {volume} {98}},\
  \bibinfo {pages} {106803} (\bibinfo {year} {2007})}\BibitemShut {NoStop}%
\bibitem [{\citenamefont {Hasan}\ and\ \citenamefont
  {Kane}(2010)}]{Hasan_colloquim_2010}%
  \BibitemOpen
  \bibfield  {author} {\bibinfo {author} {\bibfnamefont {M.~Z.}\ \bibnamefont
  {Hasan}}\ and\ \bibinfo {author} {\bibfnamefont {C.~L.}\ \bibnamefont
  {Kane}},\ }\href {https://doi.org/10.1103/RevModPhys.82.3045} {\bibfield
  {journal} {\bibinfo  {journal} {Rev. Mod. Phys.}\ }\textbf {\bibinfo {volume}
  {82}},\ \bibinfo {pages} {3045} (\bibinfo {year} {2010})}\BibitemShut
  {NoStop}%
\bibitem [{\citenamefont {Qi}\ and\ \citenamefont
  {Zhang}(2011)}]{qi_topological_2011}%
  \BibitemOpen
  \bibfield  {author} {\bibinfo {author} {\bibfnamefont {X.-L.}\ \bibnamefont
  {Qi}}\ and\ \bibinfo {author} {\bibfnamefont {S.-C.}\ \bibnamefont {Zhang}},\
  }\href {https://doi.org/10.1103/RevModPhys.83.1057} {\bibfield  {journal}
  {\bibinfo  {journal} {Rev. Mod. Phys.}\ }\textbf {\bibinfo {volume} {83}},\
  \bibinfo {pages} {1057} (\bibinfo {year} {2011})}\BibitemShut {NoStop}%
\bibitem [{\citenamefont {Kallin}\ and\ \citenamefont
  {Berlinsky}(2016)}]{kallin_chiral_2016}%
  \BibitemOpen
  \bibfield  {author} {\bibinfo {author} {\bibfnamefont {C.}~\bibnamefont
  {Kallin}}\ and\ \bibinfo {author} {\bibfnamefont {J.}~\bibnamefont
  {Berlinsky}},\ }\href {https://doi.org/10.1088/0034-4885/79/5/054502}
  {\bibfield  {journal} {\bibinfo  {journal} {Rep. Prog. Phys.}\ }\textbf
  {\bibinfo {volume} {79}},\ \bibinfo {pages} {054502} (\bibinfo {year}
  {2016})}\BibitemShut {NoStop}%
\bibitem [{\citenamefont {Sato}\ and\ \citenamefont
  {Ando}(2017)}]{sato_topological_2017}%
  \BibitemOpen
  \bibfield  {author} {\bibinfo {author} {\bibfnamefont {M.}~\bibnamefont
  {Sato}}\ and\ \bibinfo {author} {\bibfnamefont {Y.}~\bibnamefont {Ando}},\
  }\href {https://doi.org/10.1088/1361-6633/aa6ac7} {\bibfield  {journal}
  {\bibinfo  {journal} {Rep. Prog. Phys.}\ }\textbf {\bibinfo {volume} {80}},\
  \bibinfo {pages} {076501} (\bibinfo {year} {2017})}\BibitemShut {NoStop}%
\bibitem [{\citenamefont {Armitage}\ \emph {et~al.}(2018)\citenamefont
  {Armitage}, \citenamefont {Mele},\ and\ \citenamefont
  {Vishwanath}}]{armitage_weyl_2018}%
  \BibitemOpen
  \bibfield  {author} {\bibinfo {author} {\bibfnamefont {N.}~\bibnamefont
  {Armitage}}, \bibinfo {author} {\bibfnamefont {E.}~\bibnamefont {Mele}},\
  and\ \bibinfo {author} {\bibfnamefont {A.}~\bibnamefont {Vishwanath}},\
  }\href {https://doi.org/10.1103/RevModPhys.90.015001} {\bibfield  {journal}
  {\bibinfo  {journal} {Rev. Mod. Phys.}\ }\textbf {\bibinfo {volume} {90}},\
  \bibinfo {pages} {015001} (\bibinfo {year} {2018})}\BibitemShut {NoStop}%
\bibitem [{\citenamefont {Hosur}\ and\ \citenamefont
  {Qi}(2013)}]{hosur_recent_2013}%
  \BibitemOpen
  \bibfield  {author} {\bibinfo {author} {\bibfnamefont {P.}~\bibnamefont
  {Hosur}}\ and\ \bibinfo {author} {\bibfnamefont {X.}~\bibnamefont {Qi}},\
  }\href {https://doi.org/10.1016/j.crhy.2013.10.010} {\bibfield  {journal}
  {\bibinfo  {journal} {Comptes Rendus Physique}\ }\textbf {\bibinfo {volume}
  {14}},\ \bibinfo {pages} {857} (\bibinfo {year} {2013})}\BibitemShut
  {NoStop}%
\bibitem [{\citenamefont {Jia}\ \emph {et~al.}(2016)\citenamefont {Jia},
  \citenamefont {Xu},\ and\ \citenamefont {Hasan}}]{jia_weyl_2016}%
  \BibitemOpen
  \bibfield  {author} {\bibinfo {author} {\bibfnamefont {S.}~\bibnamefont
  {Jia}}, \bibinfo {author} {\bibfnamefont {S.-Y.}\ \bibnamefont {Xu}},\ and\
  \bibinfo {author} {\bibfnamefont {M.~Z.}\ \bibnamefont {Hasan}},\ }\href@noop
  {} {\bibfield  {journal} {\bibinfo  {journal} {NATURE MATERIALS}\ }\textbf
  {\bibinfo {volume} {15}},\ \bibinfo {pages} {5} (\bibinfo {year}
  {2016})}\BibitemShut {NoStop}%
\bibitem [{\citenamefont {Wehling}\ \emph {et~al.}(2014)\citenamefont
  {Wehling}, \citenamefont {Black-Schaffer},\ and\ \citenamefont
  {Balatsky}}]{wehling_dirac_2014}%
  \BibitemOpen
  \bibfield  {author} {\bibinfo {author} {\bibfnamefont {T.~O.}\ \bibnamefont
  {Wehling}}, \bibinfo {author} {\bibfnamefont {A.~M.}\ \bibnamefont
  {Black-Schaffer}},\ and\ \bibinfo {author} {\bibfnamefont {A.~V.}\
  \bibnamefont {Balatsky}},\ }\href
  {https://doi.org/10.1080/00018732.2014.927109} {\bibfield  {journal}
  {\bibinfo  {journal} {Advances in Physics}\ }\textbf {\bibinfo {volume}
  {63}},\ \bibinfo {pages} {1} (\bibinfo {year} {2014})}\BibitemShut {NoStop}%
\bibitem [{\citenamefont {Gao}\ \emph {et~al.}(2019)\citenamefont {Gao},
  \citenamefont {Venderbos}, \citenamefont {Kim},\ and\ \citenamefont
  {Rappe}}]{gao_topological_2019}%
  \BibitemOpen
  \bibfield  {author} {\bibinfo {author} {\bibfnamefont {H.}~\bibnamefont
  {Gao}}, \bibinfo {author} {\bibfnamefont {J.~W.~F.}\ \bibnamefont
  {Venderbos}}, \bibinfo {author} {\bibfnamefont {Y.}~\bibnamefont {Kim}},\
  and\ \bibinfo {author} {\bibfnamefont {A.~M.}\ \bibnamefont {Rappe}},\ }\href
  {https://doi.org/10.1146/annurev-matsci-070218-010049} {\bibfield  {journal}
  {\bibinfo  {journal} {Annu. Rev. Mater. Res.}\ }\textbf {\bibinfo {volume}
  {49}},\ \bibinfo {pages} {153} (\bibinfo {year} {2019})}\BibitemShut
  {NoStop}%
\bibitem [{\citenamefont {Vafek}\ and\ \citenamefont
  {Vishwanath}(2014)}]{vafek_dirac_2014}%
  \BibitemOpen
  \bibfield  {author} {\bibinfo {author} {\bibfnamefont {O.}~\bibnamefont
  {Vafek}}\ and\ \bibinfo {author} {\bibfnamefont {A.}~\bibnamefont
  {Vishwanath}},\ }\href
  {https://doi.org/10.1146/annurev-conmatphys-031113-133841} {\bibfield
  {journal} {\bibinfo  {journal} {Annu. Rev. Condens. Matter Phys.}\ }\textbf
  {\bibinfo {volume} {5}},\ \bibinfo {pages} {83} (\bibinfo {year}
  {2014})}\BibitemShut {NoStop}%
\bibitem [{\citenamefont {Weyl}(1929)}]{weyl_elektron_1929}%
  \BibitemOpen
  \bibfield  {author} {\bibinfo {author} {\bibfnamefont {H.}~\bibnamefont
  {Weyl}},\ }\href {https://doi.org/10.1007/BF01339504} {\bibfield  {journal}
  {\bibinfo  {journal} {Z. Physik}\ }\textbf {\bibinfo {volume} {56}},\
  \bibinfo {pages} {330} (\bibinfo {year} {1929})}\BibitemShut {NoStop}%
\bibitem [{\citenamefont {Herring}(1937)}]{herring_accidental_1937}%
  \BibitemOpen
  \bibfield  {author} {\bibinfo {author} {\bibfnamefont {C.}~\bibnamefont
  {Herring}},\ }\href {https://doi.org/10.1103/PhysRev.52.365} {\bibfield
  {journal} {\bibinfo  {journal} {Phys. Rev.}\ }\textbf {\bibinfo {volume}
  {52}},\ \bibinfo {pages} {365} (\bibinfo {year} {1937})}\BibitemShut
  {NoStop}%
\bibitem [{\citenamefont {Burkov}\ \emph {et~al.}(2011)\citenamefont {Burkov},
  \citenamefont {Hook},\ and\ \citenamefont
  {Balents}}]{burkov_topological_2011}%
  \BibitemOpen
  \bibfield  {author} {\bibinfo {author} {\bibfnamefont {A.~A.}\ \bibnamefont
  {Burkov}}, \bibinfo {author} {\bibfnamefont {M.~D.}\ \bibnamefont {Hook}},\
  and\ \bibinfo {author} {\bibfnamefont {L.}~\bibnamefont {Balents}},\ }\href
  {https://doi.org/10.1103/PhysRevB.84.235126} {\bibfield  {journal} {\bibinfo
  {journal} {Phys. Rev. B}\ }\textbf {\bibinfo {volume} {84}},\ \bibinfo
  {pages} {235126} (\bibinfo {year} {2011})}\BibitemShut {NoStop}%
\bibitem [{\citenamefont {Halász}\ and\ \citenamefont
  {Balents}(2012)}]{halasz_time-reversal_2012}%
  \BibitemOpen
  \bibfield  {author} {\bibinfo {author} {\bibfnamefont {G.~B.}\ \bibnamefont
  {Halász}}\ and\ \bibinfo {author} {\bibfnamefont {L.}~\bibnamefont
  {Balents}},\ }\href {https://doi.org/10.1103/PhysRevB.85.035103} {\bibfield
  {journal} {\bibinfo  {journal} {Phys. Rev. B}\ }\textbf {\bibinfo {volume}
  {85}},\ \bibinfo {pages} {035103} (\bibinfo {year} {2012})}\BibitemShut
  {NoStop}%
\bibitem [{\citenamefont {Delplace}\ \emph {et~al.}(2012)\citenamefont
  {Delplace}, \citenamefont {Li},\ and\ \citenamefont
  {Carpentier}}]{delplace_topological_2012}%
  \BibitemOpen
  \bibfield  {author} {\bibinfo {author} {\bibfnamefont {P.}~\bibnamefont
  {Delplace}}, \bibinfo {author} {\bibfnamefont {J.}~\bibnamefont {Li}},\ and\
  \bibinfo {author} {\bibfnamefont {D.}~\bibnamefont {Carpentier}},\ }\href
  {https://doi.org/10.1209/0295-5075/97/67004} {\bibfield  {journal} {\bibinfo
  {journal} {EPL}\ }\textbf {\bibinfo {volume} {97}},\ \bibinfo {pages} {67004}
  (\bibinfo {year} {2012})}\BibitemShut {NoStop}%
\bibitem [{\citenamefont {Pesin}\ and\ \citenamefont
  {Balents}(2010)}]{pesin_mott_2010}%
  \BibitemOpen
  \bibfield  {author} {\bibinfo {author} {\bibfnamefont {D.}~\bibnamefont
  {Pesin}}\ and\ \bibinfo {author} {\bibfnamefont {L.}~\bibnamefont
  {Balents}},\ }\href {https://doi.org/10.1038/nphys1606} {\bibfield  {journal}
  {\bibinfo  {journal} {Nature Phys}\ }\textbf {\bibinfo {volume} {6}},\
  \bibinfo {pages} {376} (\bibinfo {year} {2010})}\BibitemShut {NoStop}%
\bibitem [{\citenamefont {Wan}\ \emph {et~al.}(2011)\citenamefont {Wan},
  \citenamefont {Turner}, \citenamefont {Vishwanath},\ and\ \citenamefont
  {Savrasov}}]{wan_topological_2011}%
  \BibitemOpen
  \bibfield  {author} {\bibinfo {author} {\bibfnamefont {X.}~\bibnamefont
  {Wan}}, \bibinfo {author} {\bibfnamefont {A.~M.}\ \bibnamefont {Turner}},
  \bibinfo {author} {\bibfnamefont {A.}~\bibnamefont {Vishwanath}},\ and\
  \bibinfo {author} {\bibfnamefont {S.~Y.}\ \bibnamefont {Savrasov}},\ }\href
  {https://doi.org/10.1103/PhysRevB.83.205101} {\bibfield  {journal} {\bibinfo
  {journal} {Phys. Rev. B}\ }\textbf {\bibinfo {volume} {83}},\ \bibinfo
  {pages} {205101} (\bibinfo {year} {2011})}\BibitemShut {NoStop}%
\bibitem [{\citenamefont {Yang}\ \emph {et~al.}(2011)\citenamefont {Yang},
  \citenamefont {Lu},\ and\ \citenamefont {Ran}}]{yang_quantum_2011}%
  \BibitemOpen
  \bibfield  {author} {\bibinfo {author} {\bibfnamefont {K.-Y.}\ \bibnamefont
  {Yang}}, \bibinfo {author} {\bibfnamefont {Y.-M.}\ \bibnamefont {Lu}},\ and\
  \bibinfo {author} {\bibfnamefont {Y.}~\bibnamefont {Ran}},\ }\href
  {https://doi.org/10.1103/PhysRevB.84.075129} {\bibfield  {journal} {\bibinfo
  {journal} {Phys. Rev. B}\ }\textbf {\bibinfo {volume} {84}},\ \bibinfo
  {pages} {075129} (\bibinfo {year} {2011})}\BibitemShut {NoStop}%
\bibitem [{\citenamefont {{Shuichi
  Murakami}}(2007)}]{shuichi_murakami_phase_2007}%
  \BibitemOpen
  \bibfield  {author} {\bibinfo {author} {\bibnamefont {{Shuichi Murakami}}},\
  }\href {https://doi.org/10.1088/1367-2630/9/9/356} {\bibfield  {journal}
  {\bibinfo  {journal} {New J. Phys.}\ }\textbf {\bibinfo {volume} {9}},\
  \bibinfo {pages} {356} (\bibinfo {year} {2007})}\BibitemShut {NoStop}%
\bibitem [{\citenamefont {Hosur}\ \emph {et~al.}(2012)\citenamefont {Hosur},
  \citenamefont {Parameswaran},\ and\ \citenamefont
  {Vishwanath}}]{hosur_charge_2012}%
  \BibitemOpen
  \bibfield  {author} {\bibinfo {author} {\bibfnamefont {P.}~\bibnamefont
  {Hosur}}, \bibinfo {author} {\bibfnamefont {S.~A.}\ \bibnamefont
  {Parameswaran}},\ and\ \bibinfo {author} {\bibfnamefont {A.}~\bibnamefont
  {Vishwanath}},\ }\href {https://doi.org/10.1103/PhysRevLett.108.046602}
  {\bibfield  {journal} {\bibinfo  {journal} {Phys. Rev. Lett.}\ }\textbf
  {\bibinfo {volume} {108}},\ \bibinfo {pages} {046602} (\bibinfo {year}
  {2012})}\BibitemShut {NoStop}%
\bibitem [{\citenamefont {Burkov}\ and\ \citenamefont
  {Balents}(2011)}]{burkov_weyl_2011-1}%
  \BibitemOpen
  \bibfield  {author} {\bibinfo {author} {\bibfnamefont {A.~A.}\ \bibnamefont
  {Burkov}}\ and\ \bibinfo {author} {\bibfnamefont {L.}~\bibnamefont
  {Balents}},\ }\href {https://doi.org/10.1103/PhysRevLett.107.127205}
  {\bibfield  {journal} {\bibinfo  {journal} {Phys. Rev. Lett.}\ }\textbf
  {\bibinfo {volume} {107}},\ \bibinfo {pages} {127205} (\bibinfo {year}
  {2011})}\BibitemShut {NoStop}%
\bibitem [{\citenamefont {Zyuzin}\ \emph {et~al.}(2012)\citenamefont {Zyuzin},
  \citenamefont {Wu},\ and\ \citenamefont {Burkov}}]{zyuzin_weyl_2012}%
  \BibitemOpen
  \bibfield  {author} {\bibinfo {author} {\bibfnamefont {A.~A.}\ \bibnamefont
  {Zyuzin}}, \bibinfo {author} {\bibfnamefont {S.}~\bibnamefont {Wu}},\ and\
  \bibinfo {author} {\bibfnamefont {A.~A.}\ \bibnamefont {Burkov}},\ }\href
  {https://doi.org/10.1103/PhysRevB.85.165110} {\bibfield  {journal} {\bibinfo
  {journal} {Phys. Rev. B}\ }\textbf {\bibinfo {volume} {85}},\ \bibinfo
  {pages} {165110} (\bibinfo {year} {2012})}\BibitemShut {NoStop}%
\bibitem [{\citenamefont {Zyuzin}\ and\ \citenamefont
  {Burkov}(2012)}]{zyuzin_topological_2012}%
  \BibitemOpen
  \bibfield  {author} {\bibinfo {author} {\bibfnamefont {A.~A.}\ \bibnamefont
  {Zyuzin}}\ and\ \bibinfo {author} {\bibfnamefont {A.~A.}\ \bibnamefont
  {Burkov}},\ }\href {https://doi.org/10.1103/PhysRevB.86.115133} {\bibfield
  {journal} {\bibinfo  {journal} {Phys. Rev. B}\ }\textbf {\bibinfo {volume}
  {86}},\ \bibinfo {pages} {115133} (\bibinfo {year} {2012})}\BibitemShut
  {NoStop}%
\bibitem [{\citenamefont {Goswami}\ and\ \citenamefont
  {Tewari}(2013)}]{goswami_axionic_2013}%
  \BibitemOpen
  \bibfield  {author} {\bibinfo {author} {\bibfnamefont {P.}~\bibnamefont
  {Goswami}}\ and\ \bibinfo {author} {\bibfnamefont {S.}~\bibnamefont
  {Tewari}},\ }\href {https://doi.org/10.1103/PhysRevB.88.245107} {\bibfield
  {journal} {\bibinfo  {journal} {Phys. Rev. B}\ }\textbf {\bibinfo {volume}
  {88}},\ \bibinfo {pages} {245107} (\bibinfo {year} {2013})}\BibitemShut
  {NoStop}%
\bibitem [{\citenamefont {Grushin}(2012)}]{grushin_consequences_2012}%
  \BibitemOpen
  \bibfield  {author} {\bibinfo {author} {\bibfnamefont {A.~G.}\ \bibnamefont
  {Grushin}},\ }\href {https://doi.org/10.1103/PhysRevD.86.045001} {\bibfield
  {journal} {\bibinfo  {journal} {Phys. Rev. D}\ }\textbf {\bibinfo {volume}
  {86}},\ \bibinfo {pages} {045001} (\bibinfo {year} {2012})}\BibitemShut
  {NoStop}%
\bibitem [{\citenamefont {Vazifeh}\ and\ \citenamefont
  {Franz}(2013)}]{vazifeh_electromagnetic_2013}%
  \BibitemOpen
  \bibfield  {author} {\bibinfo {author} {\bibfnamefont {M.~M.}\ \bibnamefont
  {Vazifeh}}\ and\ \bibinfo {author} {\bibfnamefont {M.}~\bibnamefont
  {Franz}},\ }\href {https://doi.org/10.1103/PhysRevLett.111.027201} {\bibfield
   {journal} {\bibinfo  {journal} {Phys. Rev. Lett.}\ }\textbf {\bibinfo
  {volume} {111}},\ \bibinfo {pages} {027201} (\bibinfo {year}
  {2013})}\BibitemShut {NoStop}%
\bibitem [{\citenamefont {Nielsen}\ and\ \citenamefont
  {Ninomiya}(1981{\natexlab{a}})}]{nielsen_absence_1981}%
  \BibitemOpen
  \bibfield  {author} {\bibinfo {author} {\bibfnamefont {H.~B.}\ \bibnamefont
  {Nielsen}}\ and\ \bibinfo {author} {\bibfnamefont {M.}~\bibnamefont
  {Ninomiya}},\ }\href {https://doi.org/10.1016/0550-3213(81)90361-8}
  {\bibfield  {journal} {\bibinfo  {journal} {Nuclear Physics B}\ }\textbf
  {\bibinfo {volume} {185}},\ \bibinfo {pages} {20} (\bibinfo {year}
  {1981}{\natexlab{a}})}\BibitemShut {NoStop}%
\bibitem [{\citenamefont {Nielsen}\ and\ \citenamefont
  {Ninomiya}(1981{\natexlab{b}})}]{nielsen_absence_1981(2)}%
  \BibitemOpen
  \bibfield  {author} {\bibinfo {author} {\bibfnamefont {H.~B.}\ \bibnamefont
  {Nielsen}}\ and\ \bibinfo {author} {\bibfnamefont {M.}~\bibnamefont
  {Ninomiya}},\ }\href {https://doi.org/10.1016/0550-3213(81)90524-1}
  {\bibfield  {journal} {\bibinfo  {journal} {Nuclear Physics B}\ }\textbf
  {\bibinfo {volume} {193}},\ \bibinfo {pages} {173} (\bibinfo {year}
  {1981}{\natexlab{b}})}\BibitemShut {NoStop}%
\bibitem [{\citenamefont {Volovik}(2007)}]{volovik_quantum_2007}%
  \BibitemOpen
  \bibfield  {author} {\bibinfo {author} {\bibfnamefont {G.~E.}\ \bibnamefont
  {Volovik}},\ }in\ \href {https://doi.org/10.1007/3-540-70859-6_3} {\emph
  {\bibinfo {booktitle} {Quantum {Analogues}: {From} {Phase} {Transitions} to
  {Black} {Holes} and {Cosmology}}}},\ \bibinfo {series and number} {Lecture
  {Notes} in {Physics}},\ \bibinfo {editor} {edited by\ \bibinfo {editor}
  {\bibfnamefont {W.~G.}\ \bibnamefont {Unruh}}\ and\ \bibinfo {editor}
  {\bibfnamefont {R.}~\bibnamefont {Schützhold}}}\ (\bibinfo  {publisher}
  {Springer},\ \bibinfo {address} {Berlin, Heidelberg},\ \bibinfo {year}
  {2007})\ pp.\ \bibinfo {pages} {31--73}\BibitemShut {NoStop}%
\bibitem [{\citenamefont {Klinkhamer}\ and\ \citenamefont
  {Volovik}(2005)}]{klinkhamer_emergent_2005}%
  \BibitemOpen
  \bibfield  {author} {\bibinfo {author} {\bibfnamefont {F.~R.}\ \bibnamefont
  {Klinkhamer}}\ and\ \bibinfo {author} {\bibfnamefont {G.~E.}\ \bibnamefont
  {Volovik}},\ }\href {https://doi.org/10.1142/S0217751X05020902} {\bibfield
  {journal} {\bibinfo  {journal} {Int. J. Mod. Phys. A}\ }\textbf {\bibinfo
  {volume} {20}},\ \bibinfo {pages} {2795} (\bibinfo {year}
  {2005})}\BibitemShut {NoStop}%
\bibitem [{\citenamefont {Xu}\ \emph {et~al.}(2015)\citenamefont {Xu},
  \citenamefont {Belopolski}, \citenamefont {Alidoust}, \citenamefont
  {Neupane}, \citenamefont {Bian}, \citenamefont {Zhang}, \citenamefont
  {Sankar}, \citenamefont {Chang}, \citenamefont {Yuan}, \citenamefont {Lee},
  \citenamefont {Huang}, \citenamefont {Zheng}, \citenamefont {Ma},
  \citenamefont {Sanchez}, \citenamefont {Wang}, \citenamefont {Bansil},
  \citenamefont {Chou}, \citenamefont {Shibayev}, \citenamefont {Lin},
  \citenamefont {Jia},\ and\ \citenamefont {Hasan}}]{xu_discovery_2015}%
  \BibitemOpen
  \bibfield  {author} {\bibinfo {author} {\bibfnamefont {S.-Y.}\ \bibnamefont
  {Xu}}, \bibinfo {author} {\bibfnamefont {I.}~\bibnamefont {Belopolski}},
  \bibinfo {author} {\bibfnamefont {N.}~\bibnamefont {Alidoust}}, \bibinfo
  {author} {\bibfnamefont {M.}~\bibnamefont {Neupane}}, \bibinfo {author}
  {\bibfnamefont {G.}~\bibnamefont {Bian}}, \bibinfo {author} {\bibfnamefont
  {C.}~\bibnamefont {Zhang}}, \bibinfo {author} {\bibfnamefont
  {R.}~\bibnamefont {Sankar}}, \bibinfo {author} {\bibfnamefont
  {G.}~\bibnamefont {Chang}}, \bibinfo {author} {\bibfnamefont
  {Z.}~\bibnamefont {Yuan}}, \bibinfo {author} {\bibfnamefont {C.-C.}\
  \bibnamefont {Lee}}, \bibinfo {author} {\bibfnamefont {S.-M.}\ \bibnamefont
  {Huang}}, \bibinfo {author} {\bibfnamefont {H.}~\bibnamefont {Zheng}},
  \bibinfo {author} {\bibfnamefont {J.}~\bibnamefont {Ma}}, \bibinfo {author}
  {\bibfnamefont {D.~S.}\ \bibnamefont {Sanchez}}, \bibinfo {author}
  {\bibfnamefont {B.}~\bibnamefont {Wang}}, \bibinfo {author} {\bibfnamefont
  {A.}~\bibnamefont {Bansil}}, \bibinfo {author} {\bibfnamefont
  {F.}~\bibnamefont {Chou}}, \bibinfo {author} {\bibfnamefont {P.~P.}\
  \bibnamefont {Shibayev}}, \bibinfo {author} {\bibfnamefont {H.}~\bibnamefont
  {Lin}}, \bibinfo {author} {\bibfnamefont {S.}~\bibnamefont {Jia}},\ and\
  \bibinfo {author} {\bibfnamefont {M.~Z.}\ \bibnamefont {Hasan}},\ }\href
  {https://doi.org/10.1126/science.aaa9297} {\bibfield  {journal} {\bibinfo
  {journal} {Science}\ }\textbf {\bibinfo {volume} {349}},\ \bibinfo {pages}
  {613} (\bibinfo {year} {2015})}\BibitemShut {NoStop}%
\bibitem [{\citenamefont {Lu}\ \emph {et~al.}(2015{\natexlab{a}})\citenamefont
  {Lu}, \citenamefont {Wang}, \citenamefont {Ye}, \citenamefont {Ran},
  \citenamefont {Fu}, \citenamefont {Joannopoulos},\ and\ \citenamefont
  {Solja i}}]{lu_experimental_2015}%
  \BibitemOpen
  \bibfield  {author} {\bibinfo {author} {\bibfnamefont {L.}~\bibnamefont
  {Lu}}, \bibinfo {author} {\bibfnamefont {Z.}~\bibnamefont {Wang}}, \bibinfo
  {author} {\bibfnamefont {D.}~\bibnamefont {Ye}}, \bibinfo {author}
  {\bibfnamefont {L.}~\bibnamefont {Ran}}, \bibinfo {author} {\bibfnamefont
  {L.}~\bibnamefont {Fu}}, \bibinfo {author} {\bibfnamefont {J.~D.}\
  \bibnamefont {Joannopoulos}},\ and\ \bibinfo {author} {\bibfnamefont
  {M.}~\bibnamefont {Solja i}},\ }\href
  {https://doi.org/10.1126/science.aaa9273} {\bibfield  {journal} {\bibinfo
  {journal} {Science}\ }\textbf {\bibinfo {volume} {349}},\ \bibinfo {pages}
  {622} (\bibinfo {year} {2015}{\natexlab{a}})}\BibitemShut {NoStop}%
\bibitem [{\citenamefont {Belopolski}\ \emph {et~al.}(2016)\citenamefont
  {Belopolski}, \citenamefont {Xu}, \citenamefont {Sanchez}, \citenamefont
  {Chang}, \citenamefont {Guo}, \citenamefont {Neupane}, \citenamefont {Zheng},
  \citenamefont {Lee}, \citenamefont {Huang}, \citenamefont {Bian},
  \citenamefont {Alidoust}, \citenamefont {Chang}, \citenamefont {Wang},
  \citenamefont {Zhang}, \citenamefont {Bansil}, \citenamefont {Jeng},
  \citenamefont {Lin}, \citenamefont {Jia},\ and\ \citenamefont
  {Hasan}}]{belopolski_criteria_2016}%
  \BibitemOpen
  \bibfield  {author} {\bibinfo {author} {\bibfnamefont {I.}~\bibnamefont
  {Belopolski}}, \bibinfo {author} {\bibfnamefont {S.-Y.}\ \bibnamefont {Xu}},
  \bibinfo {author} {\bibfnamefont {D.}~\bibnamefont {Sanchez}}, \bibinfo
  {author} {\bibfnamefont {G.}~\bibnamefont {Chang}}, \bibinfo {author}
  {\bibfnamefont {C.}~\bibnamefont {Guo}}, \bibinfo {author} {\bibfnamefont
  {M.}~\bibnamefont {Neupane}}, \bibinfo {author} {\bibfnamefont
  {H.}~\bibnamefont {Zheng}}, \bibinfo {author} {\bibfnamefont {C.-C.}\
  \bibnamefont {Lee}}, \bibinfo {author} {\bibfnamefont {S.-M.}\ \bibnamefont
  {Huang}}, \bibinfo {author} {\bibfnamefont {G.}~\bibnamefont {Bian}},
  \bibinfo {author} {\bibfnamefont {N.}~\bibnamefont {Alidoust}}, \bibinfo
  {author} {\bibfnamefont {T.-R.}\ \bibnamefont {Chang}}, \bibinfo {author}
  {\bibfnamefont {B.}~\bibnamefont {Wang}}, \bibinfo {author} {\bibfnamefont
  {X.}~\bibnamefont {Zhang}}, \bibinfo {author} {\bibfnamefont
  {A.}~\bibnamefont {Bansil}}, \bibinfo {author} {\bibfnamefont {H.-T.}\
  \bibnamefont {Jeng}}, \bibinfo {author} {\bibfnamefont {H.}~\bibnamefont
  {Lin}}, \bibinfo {author} {\bibfnamefont {S.}~\bibnamefont {Jia}},\ and\
  \bibinfo {author} {\bibfnamefont {M.~Z.}\ \bibnamefont {Hasan}},\ }\href
  {https://doi.org/10.1103/PhysRevLett.116.066802} {\bibfield  {journal}
  {\bibinfo  {journal} {Phys. Rev. Lett.}\ }\textbf {\bibinfo {volume} {116}},\
  \bibinfo {pages} {066802} (\bibinfo {year} {2016})}\BibitemShut {NoStop}%
\bibitem [{\citenamefont {Liu}\ \emph {et~al.}(2013)\citenamefont {Liu},
  \citenamefont {Ye},\ and\ \citenamefont {Qi}}]{liu_chiral_2013}%
  \BibitemOpen
  \bibfield  {author} {\bibinfo {author} {\bibfnamefont {C.-X.}\ \bibnamefont
  {Liu}}, \bibinfo {author} {\bibfnamefont {P.}~\bibnamefont {Ye}},\ and\
  \bibinfo {author} {\bibfnamefont {X.-L.}\ \bibnamefont {Qi}},\ }\href
  {https://doi.org/10.1103/PhysRevB.87.235306} {\bibfield  {journal} {\bibinfo
  {journal} {Phys. Rev. B}\ }\textbf {\bibinfo {volume} {87}},\ \bibinfo
  {pages} {235306} (\bibinfo {year} {2013})}\BibitemShut {NoStop}%
\bibitem [{\citenamefont {Fukushima}\ \emph {et~al.}(2008)\citenamefont
  {Fukushima}, \citenamefont {Kharzeev},\ and\ \citenamefont
  {Warringa}}]{fukushima_chiral_2008}%
  \BibitemOpen
  \bibfield  {author} {\bibinfo {author} {\bibfnamefont {K.}~\bibnamefont
  {Fukushima}}, \bibinfo {author} {\bibfnamefont {D.~E.}\ \bibnamefont
  {Kharzeev}},\ and\ \bibinfo {author} {\bibfnamefont {H.~J.}\ \bibnamefont
  {Warringa}},\ }\href {https://doi.org/10.1103/PhysRevD.78.074033} {\bibfield
  {journal} {\bibinfo  {journal} {Phys. Rev. D}\ }\textbf {\bibinfo {volume}
  {78}},\ \bibinfo {pages} {074033} (\bibinfo {year} {2008})}\BibitemShut
  {NoStop}%
\bibitem [{\citenamefont {Son}\ and\ \citenamefont
  {Spivak}(2013)}]{son_chiral_2013}%
  \BibitemOpen
  \bibfield  {author} {\bibinfo {author} {\bibfnamefont {D.~T.}\ \bibnamefont
  {Son}}\ and\ \bibinfo {author} {\bibfnamefont {B.~Z.}\ \bibnamefont
  {Spivak}},\ }\href {https://doi.org/10.1103/PhysRevB.88.104412} {\bibfield
  {journal} {\bibinfo  {journal} {Phys. Rev. B}\ }\textbf {\bibinfo {volume}
  {88}},\ \bibinfo {pages} {104412} (\bibinfo {year} {2013})}\BibitemShut
  {NoStop}%
\bibitem [{\citenamefont {Kharzeev}\ and\ \citenamefont
  {Warringa}(2009)}]{kharzeev_chiral_2009}%
  \BibitemOpen
  \bibfield  {author} {\bibinfo {author} {\bibfnamefont {D.~E.}\ \bibnamefont
  {Kharzeev}}\ and\ \bibinfo {author} {\bibfnamefont {H.~J.}\ \bibnamefont
  {Warringa}},\ }\href {https://doi.org/10.1103/PhysRevD.80.034028} {\bibfield
  {journal} {\bibinfo  {journal} {Phys. Rev. D}\ }\textbf {\bibinfo {volume}
  {80}},\ \bibinfo {pages} {034028} (\bibinfo {year} {2009})}\BibitemShut
  {NoStop}%
\bibitem [{\citenamefont {Huang}\ \emph {et~al.}(2015)\citenamefont {Huang},
  \citenamefont {Zhao}, \citenamefont {Long}, \citenamefont {Wang},
  \citenamefont {Chen}, \citenamefont {Yang}, \citenamefont {Liang},
  \citenamefont {Xue}, \citenamefont {Weng}, \citenamefont {Fang},
  \citenamefont {Dai},\ and\ \citenamefont {Chen}}]{huang_observation_2015}%
  \BibitemOpen
  \bibfield  {author} {\bibinfo {author} {\bibfnamefont {X.}~\bibnamefont
  {Huang}}, \bibinfo {author} {\bibfnamefont {L.}~\bibnamefont {Zhao}},
  \bibinfo {author} {\bibfnamefont {Y.}~\bibnamefont {Long}}, \bibinfo {author}
  {\bibfnamefont {P.}~\bibnamefont {Wang}}, \bibinfo {author} {\bibfnamefont
  {D.}~\bibnamefont {Chen}}, \bibinfo {author} {\bibfnamefont {Z.}~\bibnamefont
  {Yang}}, \bibinfo {author} {\bibfnamefont {H.}~\bibnamefont {Liang}},
  \bibinfo {author} {\bibfnamefont {M.}~\bibnamefont {Xue}}, \bibinfo {author}
  {\bibfnamefont {H.}~\bibnamefont {Weng}}, \bibinfo {author} {\bibfnamefont
  {Z.}~\bibnamefont {Fang}}, \bibinfo {author} {\bibfnamefont {X.}~\bibnamefont
  {Dai}},\ and\ \bibinfo {author} {\bibfnamefont {G.}~\bibnamefont {Chen}},\
  }\href {https://doi.org/10.1103/PhysRevX.5.031023} {\bibfield  {journal}
  {\bibinfo  {journal} {Phys. Rev. X}\ }\textbf {\bibinfo {volume} {5}},\
  \bibinfo {pages} {031023} (\bibinfo {year} {2015})}\BibitemShut {NoStop}%
\bibitem [{\citenamefont {Zhang}\ \emph {et~al.}(2016)\citenamefont {Zhang},
  \citenamefont {Xu}, \citenamefont {Belopolski}, \citenamefont {Yuan},
  \citenamefont {Lin}, \citenamefont {Tong}, \citenamefont {Bian},
  \citenamefont {Alidoust}, \citenamefont {Lee}, \citenamefont {Huang},
  \citenamefont {Chang}, \citenamefont {Chang}, \citenamefont {Hsu},
  \citenamefont {Jeng}, \citenamefont {Neupane}, \citenamefont {Sanchez},
  \citenamefont {Zheng}, \citenamefont {Wang}, \citenamefont {Lin},
  \citenamefont {Zhang}, \citenamefont {Lu}, \citenamefont {Shen},
  \citenamefont {Neupert}, \citenamefont {Zahid~Hasan},\ and\ \citenamefont
  {Jia}}]{zhang_signatures_2016}%
  \BibitemOpen
  \bibfield  {author} {\bibinfo {author} {\bibfnamefont {C.-L.}\ \bibnamefont
  {Zhang}}, \bibinfo {author} {\bibfnamefont {S.-Y.}\ \bibnamefont {Xu}},
  \bibinfo {author} {\bibfnamefont {I.}~\bibnamefont {Belopolski}}, \bibinfo
  {author} {\bibfnamefont {Z.}~\bibnamefont {Yuan}}, \bibinfo {author}
  {\bibfnamefont {Z.}~\bibnamefont {Lin}}, \bibinfo {author} {\bibfnamefont
  {B.}~\bibnamefont {Tong}}, \bibinfo {author} {\bibfnamefont {G.}~\bibnamefont
  {Bian}}, \bibinfo {author} {\bibfnamefont {N.}~\bibnamefont {Alidoust}},
  \bibinfo {author} {\bibfnamefont {C.-C.}\ \bibnamefont {Lee}}, \bibinfo
  {author} {\bibfnamefont {S.-M.}\ \bibnamefont {Huang}}, \bibinfo {author}
  {\bibfnamefont {T.-R.}\ \bibnamefont {Chang}}, \bibinfo {author}
  {\bibfnamefont {G.}~\bibnamefont {Chang}}, \bibinfo {author} {\bibfnamefont
  {C.-H.}\ \bibnamefont {Hsu}}, \bibinfo {author} {\bibfnamefont {H.-T.}\
  \bibnamefont {Jeng}}, \bibinfo {author} {\bibfnamefont {M.}~\bibnamefont
  {Neupane}}, \bibinfo {author} {\bibfnamefont {D.~S.}\ \bibnamefont
  {Sanchez}}, \bibinfo {author} {\bibfnamefont {H.}~\bibnamefont {Zheng}},
  \bibinfo {author} {\bibfnamefont {J.}~\bibnamefont {Wang}}, \bibinfo {author}
  {\bibfnamefont {H.}~\bibnamefont {Lin}}, \bibinfo {author} {\bibfnamefont
  {C.}~\bibnamefont {Zhang}}, \bibinfo {author} {\bibfnamefont {H.-Z.}\
  \bibnamefont {Lu}}, \bibinfo {author} {\bibfnamefont {S.-Q.}\ \bibnamefont
  {Shen}}, \bibinfo {author} {\bibfnamefont {T.}~\bibnamefont {Neupert}},
  \bibinfo {author} {\bibfnamefont {M.}~\bibnamefont {Zahid~Hasan}},\ and\
  \bibinfo {author} {\bibfnamefont {S.}~\bibnamefont {Jia}},\ }\href
  {https://doi.org/10.1038/ncomms10735} {\bibfield  {journal} {\bibinfo
  {journal} {Nat Commun}\ }\textbf {\bibinfo {volume} {7}},\ \bibinfo {pages}
  {10735} (\bibinfo {year} {2016})}\BibitemShut {NoStop}%
\bibitem [{\citenamefont {Zaanen}\ \emph {et~al.}(2015)\citenamefont {Zaanen},
  \citenamefont {Sun}, \citenamefont {Liu},\ and\ \citenamefont
  {Schalm}}]{Zaanen:2015oix}%
  \BibitemOpen
  \bibfield  {author} {\bibinfo {author} {\bibfnamefont {J.}~\bibnamefont
  {Zaanen}}, \bibinfo {author} {\bibfnamefont {Y.-W.}\ \bibnamefont {Sun}},
  \bibinfo {author} {\bibfnamefont {Y.}~\bibnamefont {Liu}},\ and\ \bibinfo
  {author} {\bibfnamefont {K.}~\bibnamefont {Schalm}},\ }\href@noop {} {\emph
  {\bibinfo {title} {{Holographic Duality in Condensed Matter Physics}}}}\
  (\bibinfo  {publisher} {Cambridge Univ. Press},\ \bibinfo {year}
  {2015})\BibitemShut {NoStop}%
\bibitem [{\citenamefont {Hartnoll}\ \emph {et~al.}(2018)\citenamefont
  {Hartnoll}, \citenamefont {Lucas},\ and\ \citenamefont
  {Sachdev}}]{hartnoll_holographic_2018}%
  \BibitemOpen
  \bibfield  {author} {\bibinfo {author} {\bibfnamefont {S.~A.}\ \bibnamefont
  {Hartnoll}}, \bibinfo {author} {\bibfnamefont {A.}~\bibnamefont {Lucas}},\
  and\ \bibinfo {author} {\bibfnamefont {S.}~\bibnamefont {Sachdev}},\
  }\href@noop {} {\emph {\bibinfo {title} {Holographic {Quantum} {Matter}}}},\
  \bibinfo {edition} {illustrated edição}\ ed.\ (\bibinfo  {publisher} {The
  MIT Press},\ \bibinfo {address} {Cambridge, Massachusetts ; London,
  England},\ \bibinfo {year} {2018})\BibitemShut {NoStop}%
\bibitem [{\citenamefont {Nastase}(2017)}]{Nastase:2017cxp}%
  \BibitemOpen
  \bibfield  {author} {\bibinfo {author} {\bibfnamefont {H.}~\bibnamefont
  {Nastase}},\ }\href {https://doi.org/10.1017/9781316847978} {\emph {\bibinfo
  {title} {{String Theory Methods for Condensed Matter Physics}}}}\ (\bibinfo
  {publisher} {Cambridge University Press},\ \bibinfo {year}
  {2017})\BibitemShut {NoStop}%
\bibitem [{\citenamefont {Landsteiner}\ and\ \citenamefont
  {Liu}(2016)}]{Landsteiner:2015lsa}%
  \BibitemOpen
  \bibfield  {author} {\bibinfo {author} {\bibfnamefont {K.}~\bibnamefont
  {Landsteiner}}\ and\ \bibinfo {author} {\bibfnamefont {Y.}~\bibnamefont
  {Liu}},\ }\href {https://doi.org/10.1016/j.physletb.2015.12.052} {\bibfield
  {journal} {\bibinfo  {journal} {Phys. Lett. B}\ }\textbf {\bibinfo {volume}
  {753}},\ \bibinfo {pages} {453} (\bibinfo {year} {2016})},\ \Eprint
  {https://arxiv.org/abs/1505.04772} {arXiv:1505.04772 [hep-th]} \BibitemShut
  {NoStop}%
\bibitem [{\citenamefont {Landsteiner}\ \emph
  {et~al.}(2016{\natexlab{a}})\citenamefont {Landsteiner}, \citenamefont
  {Liu},\ and\ \citenamefont {Sun}}]{Landsteiner:2015pdh}%
  \BibitemOpen
  \bibfield  {author} {\bibinfo {author} {\bibfnamefont {K.}~\bibnamefont
  {Landsteiner}}, \bibinfo {author} {\bibfnamefont {Y.}~\bibnamefont {Liu}},\
  and\ \bibinfo {author} {\bibfnamefont {Y.-W.}\ \bibnamefont {Sun}},\ }\href
  {https://doi.org/10.1103/PhysRevLett.116.081602} {\bibfield  {journal}
  {\bibinfo  {journal} {Phys. Rev. Lett.}\ }\textbf {\bibinfo {volume} {116}},\
  \bibinfo {pages} {081602} (\bibinfo {year} {2016}{\natexlab{a}})},\ \Eprint
  {https://arxiv.org/abs/1511.05505} {arXiv:1511.05505 [hep-th]} \BibitemShut
  {NoStop}%
\bibitem [{\citenamefont {Ammon}\ \emph {et~al.}(2017)\citenamefont {Ammon},
  \citenamefont {Heinrich}, \citenamefont {Jim\'enez-Alba},\ and\ \citenamefont
  {Moeckel}}]{Ammon:2016mwa}%
  \BibitemOpen
  \bibfield  {author} {\bibinfo {author} {\bibfnamefont {M.}~\bibnamefont
  {Ammon}}, \bibinfo {author} {\bibfnamefont {M.}~\bibnamefont {Heinrich}},
  \bibinfo {author} {\bibfnamefont {A.}~\bibnamefont {Jim\'enez-Alba}},\ and\
  \bibinfo {author} {\bibfnamefont {S.}~\bibnamefont {Moeckel}},\ }\href
  {https://doi.org/10.1103/PhysRevLett.118.201601} {\bibfield  {journal}
  {\bibinfo  {journal} {Phys. Rev. Lett.}\ }\textbf {\bibinfo {volume} {118}},\
  \bibinfo {pages} {201601} (\bibinfo {year} {2017})},\ \Eprint
  {https://arxiv.org/abs/1612.00836} {arXiv:1612.00836 [hep-th]} \BibitemShut
  {NoStop}%
\bibitem [{\citenamefont {Landsteiner}\ \emph
  {et~al.}(2016{\natexlab{b}})\citenamefont {Landsteiner}, \citenamefont
  {Liu},\ and\ \citenamefont {Sun}}]{Landsteiner:2016stv}%
  \BibitemOpen
  \bibfield  {author} {\bibinfo {author} {\bibfnamefont {K.}~\bibnamefont
  {Landsteiner}}, \bibinfo {author} {\bibfnamefont {Y.}~\bibnamefont {Liu}},\
  and\ \bibinfo {author} {\bibfnamefont {Y.-W.}\ \bibnamefont {Sun}},\ }\href
  {https://doi.org/10.1103/PhysRevLett.117.081604} {\bibfield  {journal}
  {\bibinfo  {journal} {Phys. Rev. Lett.}\ }\textbf {\bibinfo {volume} {117}},\
  \bibinfo {pages} {081604} (\bibinfo {year} {2016}{\natexlab{b}})},\ \Eprint
  {https://arxiv.org/abs/1604.01346} {arXiv:1604.01346 [hep-th]} \BibitemShut
  {NoStop}%
\bibitem [{\citenamefont {Grignani}\ \emph {et~al.}(2017)\citenamefont
  {Grignani}, \citenamefont {Marini}, \citenamefont {Pena-Benitez},\ and\
  \citenamefont {Speziali}}]{Grignani:2016wyz}%
  \BibitemOpen
  \bibfield  {author} {\bibinfo {author} {\bibfnamefont {G.}~\bibnamefont
  {Grignani}}, \bibinfo {author} {\bibfnamefont {A.}~\bibnamefont {Marini}},
  \bibinfo {author} {\bibfnamefont {F.}~\bibnamefont {Pena-Benitez}},\ and\
  \bibinfo {author} {\bibfnamefont {S.}~\bibnamefont {Speziali}},\ }\href
  {https://doi.org/10.1007/JHEP03(2017)125} {\bibfield  {journal} {\bibinfo
  {journal} {JHEP}\ }\textbf {\bibinfo {volume} {03}},\ \bibinfo {pages}
  {125}},\ \Eprint {https://arxiv.org/abs/1612.00486} {arXiv:1612.00486
  [cond-mat.str-el]} \BibitemShut {NoStop}%
\bibitem [{\citenamefont {Copetti}\ \emph {et~al.}(2017)\citenamefont
  {Copetti}, \citenamefont {Fern\'andez-Pend\'as},\ and\ \citenamefont
  {Landsteiner}}]{Copetti:2016ewq}%
  \BibitemOpen
  \bibfield  {author} {\bibinfo {author} {\bibfnamefont {C.}~\bibnamefont
  {Copetti}}, \bibinfo {author} {\bibfnamefont {J.}~\bibnamefont
  {Fern\'andez-Pend\'as}},\ and\ \bibinfo {author} {\bibfnamefont
  {K.}~\bibnamefont {Landsteiner}},\ }\href
  {https://doi.org/10.1007/JHEP02(2017)138} {\bibfield  {journal} {\bibinfo
  {journal} {JHEP}\ }\textbf {\bibinfo {volume} {02}},\ \bibinfo {pages}
  {138}},\ \Eprint {https://arxiv.org/abs/1611.08125} {arXiv:1611.08125
  [hep-th]} \BibitemShut {NoStop}%
\bibitem [{\citenamefont {Liu}\ and\ \citenamefont
  {Sun}(2018{\natexlab{a}})}]{Liu:2018djq}%
  \BibitemOpen
  \bibfield  {author} {\bibinfo {author} {\bibfnamefont {Y.}~\bibnamefont
  {Liu}}\ and\ \bibinfo {author} {\bibfnamefont {Y.-W.}\ \bibnamefont {Sun}},\
  }\href {https://doi.org/10.1007/JHEP10(2018)189} {\bibfield  {journal}
  {\bibinfo  {journal} {JHEP}\ }\textbf {\bibinfo {volume} {10}},\ \bibinfo
  {pages} {189}},\ \Eprint {https://arxiv.org/abs/1809.00513} {arXiv:1809.00513
  [hep-th]} \BibitemShut {NoStop}%
\bibitem [{\citenamefont {Zhao}(2022)}]{Zhao:2021qfo}%
  \BibitemOpen
  \bibfield  {author} {\bibinfo {author} {\bibfnamefont {J.}~\bibnamefont
  {Zhao}},\ }\href {https://doi.org/10.1140/epjc/s10052-022-10237-9} {\bibfield
   {journal} {\bibinfo  {journal} {Eur. Phys. J. C}\ }\textbf {\bibinfo
  {volume} {82}},\ \bibinfo {pages} {300} (\bibinfo {year} {2022})},\ \Eprint
  {https://arxiv.org/abs/2111.14068} {arXiv:2111.14068 [hep-th]} \BibitemShut
  {NoStop}%
\bibitem [{\citenamefont {Zhao}(2021)}]{Zhao:2021pih}%
  \BibitemOpen
  \bibfield  {author} {\bibinfo {author} {\bibfnamefont {J.}~\bibnamefont
  {Zhao}},\ }\href {https://doi.org/10.1103/PhysRevD.104.066003} {\bibfield
  {journal} {\bibinfo  {journal} {Phys. Rev. D}\ }\textbf {\bibinfo {volume}
  {104}},\ \bibinfo {pages} {066003} (\bibinfo {year} {2021})},\ \Eprint
  {https://arxiv.org/abs/2109.07215} {arXiv:2109.07215 [hep-th]} \BibitemShut
  {NoStop}%
\bibitem [{\citenamefont {Liu}\ and\ \citenamefont
  {Sun}(2018{\natexlab{b}})}]{Liu:2018bye}%
  \BibitemOpen
  \bibfield  {author} {\bibinfo {author} {\bibfnamefont {Y.}~\bibnamefont
  {Liu}}\ and\ \bibinfo {author} {\bibfnamefont {Y.-W.}\ \bibnamefont {Sun}},\
  }\href {https://doi.org/10.1007/JHEP12(2018)072} {\bibfield  {journal}
  {\bibinfo  {journal} {JHEP}\ }\textbf {\bibinfo {volume} {12}},\ \bibinfo
  {pages} {072}},\ \Eprint {https://arxiv.org/abs/1801.09357} {arXiv:1801.09357
  [hep-th]} \BibitemShut {NoStop}%
\bibitem [{\citenamefont {Liu}\ and\ \citenamefont {Wu}(2021)}]{Liu:2020ymx}%
  \BibitemOpen
  \bibfield  {author} {\bibinfo {author} {\bibfnamefont {Y.}~\bibnamefont
  {Liu}}\ and\ \bibinfo {author} {\bibfnamefont {X.-M.}\ \bibnamefont {Wu}},\
  }\href {https://doi.org/10.1007/JHEP05(2021)141} {\bibfield  {journal}
  {\bibinfo  {journal} {JHEP}\ }\textbf {\bibinfo {volume} {05}},\ \bibinfo
  {pages} {141}},\ \Eprint {https://arxiv.org/abs/2012.12602} {arXiv:2012.12602
  [hep-th]} \BibitemShut {NoStop}%
\bibitem [{\citenamefont {Rodgers}\ \emph {et~al.}(2021)\citenamefont
  {Rodgers}, \citenamefont {Mauri}, \citenamefont {G\"ursoy},\ and\
  \citenamefont {Stoof}}]{Rodgers:2021azg}%
  \BibitemOpen
  \bibfield  {author} {\bibinfo {author} {\bibfnamefont {R.}~\bibnamefont
  {Rodgers}}, \bibinfo {author} {\bibfnamefont {E.}~\bibnamefont {Mauri}},
  \bibinfo {author} {\bibfnamefont {U.}~\bibnamefont {G\"ursoy}},\ and\
  \bibinfo {author} {\bibfnamefont {H.~T.~C.}\ \bibnamefont {Stoof}},\ }\href
  {https://doi.org/10.1007/JHEP11(2021)191} {\bibfield  {journal} {\bibinfo
  {journal} {JHEP}\ }\textbf {\bibinfo {volume} {11}},\ \bibinfo {pages}
  {191}},\ \Eprint {https://arxiv.org/abs/2109.07187} {arXiv:2109.07187
  [hep-th]} \BibitemShut {NoStop}%
\bibitem [{\citenamefont {Ji}\ \emph {et~al.}(2021)\citenamefont {Ji},
  \citenamefont {Liu}, \citenamefont {Sun},\ and\ \citenamefont
  {Zhang}}]{Ji:2021aan}%
  \BibitemOpen
  \bibfield  {author} {\bibinfo {author} {\bibfnamefont {X.}~\bibnamefont
  {Ji}}, \bibinfo {author} {\bibfnamefont {Y.}~\bibnamefont {Liu}}, \bibinfo
  {author} {\bibfnamefont {Y.-W.}\ \bibnamefont {Sun}},\ and\ \bibinfo {author}
  {\bibfnamefont {Y.-L.}\ \bibnamefont {Zhang}},\ }\href
  {https://doi.org/10.1007/JHEP12(2021)066} {\bibfield  {journal} {\bibinfo
  {journal} {JHEP}\ }\textbf {\bibinfo {volume} {12}},\ \bibinfo {pages}
  {066}},\ \Eprint {https://arxiv.org/abs/2109.05993} {arXiv:2109.05993
  [hep-th]} \BibitemShut {NoStop}%
\bibitem [{\citenamefont {Landsteiner}\ \emph {et~al.}(2020)\citenamefont
  {Landsteiner}, \citenamefont {Liu},\ and\ \citenamefont
  {Sun}}]{Landsteiner:2019kxb}%
  \BibitemOpen
  \bibfield  {author} {\bibinfo {author} {\bibfnamefont {K.}~\bibnamefont
  {Landsteiner}}, \bibinfo {author} {\bibfnamefont {Y.}~\bibnamefont {Liu}},\
  and\ \bibinfo {author} {\bibfnamefont {Y.-W.}\ \bibnamefont {Sun}},\ }\href
  {https://doi.org/10.1007/s11433-019-1477-7} {\bibfield  {journal} {\bibinfo
  {journal} {Sci. China Phys. Mech. Astron.}\ }\textbf {\bibinfo {volume}
  {63}},\ \bibinfo {pages} {250001} (\bibinfo {year} {2020})},\ \Eprint
  {https://arxiv.org/abs/1911.07978} {arXiv:1911.07978 [hep-th]} \BibitemShut
  {NoStop}%
\bibitem [{\citenamefont {Gursoy}\ \emph {et~al.}(2013)\citenamefont {Gursoy},
  \citenamefont {Jacobs}, \citenamefont {Plauschinn}, \citenamefont {Stoof},\
  and\ \citenamefont {Vandoren}}]{Gursoy:2012ie}%
  \BibitemOpen
  \bibfield  {author} {\bibinfo {author} {\bibfnamefont {U.}~\bibnamefont
  {Gursoy}}, \bibinfo {author} {\bibfnamefont {V.}~\bibnamefont {Jacobs}},
  \bibinfo {author} {\bibfnamefont {E.}~\bibnamefont {Plauschinn}}, \bibinfo
  {author} {\bibfnamefont {H.}~\bibnamefont {Stoof}},\ and\ \bibinfo {author}
  {\bibfnamefont {S.}~\bibnamefont {Vandoren}},\ }\href
  {https://doi.org/10.1007/JHEP04(2013)127} {\bibfield  {journal} {\bibinfo
  {journal} {JHEP}\ }\textbf {\bibinfo {volume} {04}},\ \bibinfo {pages}
  {127}},\ \Eprint {https://arxiv.org/abs/1209.2593} {arXiv:1209.2593 [hep-th]}
  \BibitemShut {NoStop}%
\bibitem [{\citenamefont {Jacobs}\ \emph {et~al.}(2016)\citenamefont {Jacobs},
  \citenamefont {Betzios}, \citenamefont {Gursoy},\ and\ \citenamefont
  {Stoof}}]{Jacobs:2015fiv}%
  \BibitemOpen
  \bibfield  {author} {\bibinfo {author} {\bibfnamefont {V.~P.~J.}\
  \bibnamefont {Jacobs}}, \bibinfo {author} {\bibfnamefont {P.}~\bibnamefont
  {Betzios}}, \bibinfo {author} {\bibfnamefont {U.}~\bibnamefont {Gursoy}},\
  and\ \bibinfo {author} {\bibfnamefont {H.~T.~C.}\ \bibnamefont {Stoof}},\
  }\href {https://doi.org/10.1103/PhysRevB.93.195104} {\bibfield  {journal}
  {\bibinfo  {journal} {Phys. Rev. B}\ }\textbf {\bibinfo {volume} {93}},\
  \bibinfo {pages} {195104} (\bibinfo {year} {2016})},\ \Eprint
  {https://arxiv.org/abs/1512.04883} {arXiv:1512.04883 [hep-th]} \BibitemShut
  {NoStop}%
\bibitem [{\citenamefont {Hashimoto}\ \emph {et~al.}(2017)\citenamefont
  {Hashimoto}, \citenamefont {Kinoshita}, \citenamefont {Murata},\ and\
  \citenamefont {Oka}}]{Hashimoto:2016ize}%
  \BibitemOpen
  \bibfield  {author} {\bibinfo {author} {\bibfnamefont {K.}~\bibnamefont
  {Hashimoto}}, \bibinfo {author} {\bibfnamefont {S.}~\bibnamefont
  {Kinoshita}}, \bibinfo {author} {\bibfnamefont {K.}~\bibnamefont {Murata}},\
  and\ \bibinfo {author} {\bibfnamefont {T.}~\bibnamefont {Oka}},\ }\href
  {https://doi.org/10.1007/JHEP05(2017)127} {\bibfield  {journal} {\bibinfo
  {journal} {JHEP}\ }\textbf {\bibinfo {volume} {05}},\ \bibinfo {pages}
  {127}},\ \Eprint {https://arxiv.org/abs/1611.03702} {arXiv:1611.03702
  [hep-th]} \BibitemShut {NoStop}%
\bibitem [{\citenamefont {Liu}\ and\ \citenamefont {Zhao}(2018)}]{Liu:2018spp}%
  \BibitemOpen
  \bibfield  {author} {\bibinfo {author} {\bibfnamefont {Y.}~\bibnamefont
  {Liu}}\ and\ \bibinfo {author} {\bibfnamefont {J.}~\bibnamefont {Zhao}},\
  }\href {https://doi.org/10.1007/JHEP12(2018)124} {\bibfield  {journal}
  {\bibinfo  {journal} {JHEP}\ }\textbf {\bibinfo {volume} {12}},\ \bibinfo
  {pages} {124}},\ \Eprint {https://arxiv.org/abs/1809.08601} {arXiv:1809.08601
  [hep-th]} \BibitemShut {NoStop}%
\bibitem [{\citenamefont {Song}\ \emph {et~al.}(2019)\citenamefont {Song},
  \citenamefont {Rong},\ and\ \citenamefont {Sin}}]{Song:2019asj}%
  \BibitemOpen
  \bibfield  {author} {\bibinfo {author} {\bibfnamefont {G.}~\bibnamefont
  {Song}}, \bibinfo {author} {\bibfnamefont {J.}~\bibnamefont {Rong}},\ and\
  \bibinfo {author} {\bibfnamefont {S.-J.}\ \bibnamefont {Sin}},\ }\href
  {https://doi.org/10.1007/JHEP10(2019)109} {\bibfield  {journal} {\bibinfo
  {journal} {JHEP}\ }\textbf {\bibinfo {volume} {10}},\ \bibinfo {pages}
  {109}},\ \Eprint {https://arxiv.org/abs/1904.09349} {arXiv:1904.09349
  [hep-th]} \BibitemShut {NoStop}%
\bibitem [{\citenamefont {Ji}\ \emph {et~al.}(2019)\citenamefont {Ji},
  \citenamefont {Liu},\ and\ \citenamefont {Wu}}]{Ji:2019pxx}%
  \BibitemOpen
  \bibfield  {author} {\bibinfo {author} {\bibfnamefont {X.}~\bibnamefont
  {Ji}}, \bibinfo {author} {\bibfnamefont {Y.}~\bibnamefont {Liu}},\ and\
  \bibinfo {author} {\bibfnamefont {X.-M.}\ \bibnamefont {Wu}},\ }\href
  {https://doi.org/10.1103/PhysRevD.100.126013} {\bibfield  {journal} {\bibinfo
   {journal} {Phys. Rev. D}\ }\textbf {\bibinfo {volume} {100}},\ \bibinfo
  {pages} {126013} (\bibinfo {year} {2019})},\ \Eprint
  {https://arxiv.org/abs/1904.08058} {arXiv:1904.08058 [hep-th]} \BibitemShut
  {NoStop}%
\bibitem [{\citenamefont {Baggioli}\ and\ \citenamefont
  {Giataganas}(2021)}]{Baggioli:2020cld}%
  \BibitemOpen
  \bibfield  {author} {\bibinfo {author} {\bibfnamefont {M.}~\bibnamefont
  {Baggioli}}\ and\ \bibinfo {author} {\bibfnamefont {D.}~\bibnamefont
  {Giataganas}},\ }\href {https://doi.org/10.1103/PhysRevD.103.026009}
  {\bibfield  {journal} {\bibinfo  {journal} {Phys. Rev. D}\ }\textbf {\bibinfo
  {volume} {103}},\ \bibinfo {pages} {026009} (\bibinfo {year} {2021})},\
  \Eprint {https://arxiv.org/abs/2007.07273} {arXiv:2007.07273 [hep-th]}
  \BibitemShut {NoStop}%
\bibitem [{\citenamefont {Juri\v{c}i\'c}\ \emph {et~al.}(2020)\citenamefont
  {Juri\v{c}i\'c}, \citenamefont {Salazar~Landea},\ and\ \citenamefont
  {Soto-Garrido}}]{Juricic:2020sgg}%
  \BibitemOpen
  \bibfield  {author} {\bibinfo {author} {\bibfnamefont {V.}~\bibnamefont
  {Juri\v{c}i\'c}}, \bibinfo {author} {\bibfnamefont {I.}~\bibnamefont
  {Salazar~Landea}},\ and\ \bibinfo {author} {\bibfnamefont {R.}~\bibnamefont
  {Soto-Garrido}},\ }\href {https://doi.org/10.1007/JHEP07(2020)052} {\bibfield
   {journal} {\bibinfo  {journal} {JHEP}\ }\textbf {\bibinfo {volume} {07}},\
  \bibinfo {pages} {052}},\ \Eprint {https://arxiv.org/abs/2005.10387}
  {arXiv:2005.10387 [hep-th]} \BibitemShut {NoStop}%
\bibitem [{\citenamefont {Bitaghsir~Fadafan}\ \emph {et~al.}(2021)\citenamefont
  {Bitaghsir~Fadafan}, \citenamefont {O'Bannon}, \citenamefont {Rodgers},\ and\
  \citenamefont {Russell}}]{BitaghsirFadafan:2020lkh}%
  \BibitemOpen
  \bibfield  {author} {\bibinfo {author} {\bibfnamefont {K.}~\bibnamefont
  {Bitaghsir~Fadafan}}, \bibinfo {author} {\bibfnamefont {A.}~\bibnamefont
  {O'Bannon}}, \bibinfo {author} {\bibfnamefont {R.}~\bibnamefont {Rodgers}},\
  and\ \bibinfo {author} {\bibfnamefont {M.}~\bibnamefont {Russell}},\ }\href
  {https://doi.org/10.1007/JHEP04(2021)162} {\bibfield  {journal} {\bibinfo
  {journal} {JHEP}\ }\textbf {\bibinfo {volume} {04}},\ \bibinfo {pages}
  {162}},\ \Eprint {https://arxiv.org/abs/2012.11434} {arXiv:2012.11434
  [hep-th]} \BibitemShut {NoStop}%
\bibitem [{\citenamefont {Gao}\ \emph {et~al.}(2023)\citenamefont {Gao},
  \citenamefont {Liu},\ and\ \citenamefont {Lyu}}]{Gao:2023zbd}%
  \BibitemOpen
  \bibfield  {author} {\bibinfo {author} {\bibfnamefont {L.-L.}\ \bibnamefont
  {Gao}}, \bibinfo {author} {\bibfnamefont {Y.}~\bibnamefont {Liu}},\ and\
  \bibinfo {author} {\bibfnamefont {H.-D.}\ \bibnamefont {Lyu}},\ }\href@noop
  {} {\  (\bibinfo {year} {2023})},\ \Eprint {https://arxiv.org/abs/2301.01468}
  {arXiv:2301.01468 [hep-th]} \BibitemShut {NoStop}%
\bibitem [{\citenamefont {Satake}\ \emph {et~al.}(2020)\citenamefont {Satake},
  \citenamefont {Shiogai}, \citenamefont {Mazur}, \citenamefont {Kimura},
  \citenamefont {Awaji}, \citenamefont {Fujiwara}, \citenamefont {Nojima},
  \citenamefont {Nomura}, \citenamefont {Souma}, \citenamefont {Sato},
  \citenamefont {Dietl},\ and\ \citenamefont {Tsukazaki}}]{Shiogai_et_al}%
  \BibitemOpen
  \bibfield  {author} {\bibinfo {author} {\bibfnamefont {Y.}~\bibnamefont
  {Satake}}, \bibinfo {author} {\bibfnamefont {J.}~\bibnamefont {Shiogai}},
  \bibinfo {author} {\bibfnamefont {G.~P.}\ \bibnamefont {Mazur}}, \bibinfo
  {author} {\bibfnamefont {S.}~\bibnamefont {Kimura}}, \bibinfo {author}
  {\bibfnamefont {S.}~\bibnamefont {Awaji}}, \bibinfo {author} {\bibfnamefont
  {K.}~\bibnamefont {Fujiwara}}, \bibinfo {author} {\bibfnamefont
  {T.}~\bibnamefont {Nojima}}, \bibinfo {author} {\bibfnamefont
  {K.}~\bibnamefont {Nomura}}, \bibinfo {author} {\bibfnamefont
  {S.}~\bibnamefont {Souma}}, \bibinfo {author} {\bibfnamefont
  {T.}~\bibnamefont {Sato}}, \bibinfo {author} {\bibfnamefont {T.}~\bibnamefont
  {Dietl}},\ and\ \bibinfo {author} {\bibfnamefont {A.}~\bibnamefont
  {Tsukazaki}},\ }\href {https://doi.org/10.1103/PhysRevMaterials.4.044202}
  {\bibfield  {journal} {\bibinfo  {journal} {Phys. Rev. Mater.}\ }\textbf
  {\bibinfo {volume} {4}},\ \bibinfo {pages} {044202} (\bibinfo {year}
  {2020})}\BibitemShut {NoStop}%
\bibitem [{\citenamefont {Mayo}\ \emph {et~al.}(2022)\citenamefont {Mayo},
  \citenamefont {Takahashi}, \citenamefont {Bahramy}, \citenamefont {Nomoto},
  \citenamefont {Sakai},\ and\ \citenamefont {Ishiwata}}]{Mayo_et_al}%
  \BibitemOpen
  \bibfield  {author} {\bibinfo {author} {\bibfnamefont {A.~H.}\ \bibnamefont
  {Mayo}}, \bibinfo {author} {\bibfnamefont {H.}~\bibnamefont {Takahashi}},
  \bibinfo {author} {\bibfnamefont {M.~S.}\ \bibnamefont {Bahramy}}, \bibinfo
  {author} {\bibfnamefont {A.}~\bibnamefont {Nomoto}}, \bibinfo {author}
  {\bibfnamefont {H.}~\bibnamefont {Sakai}},\ and\ \bibinfo {author}
  {\bibfnamefont {S.}~\bibnamefont {Ishiwata}},\ }\href
  {https://doi.org/10.1103/PhysRevX.12.011033} {\bibfield  {journal} {\bibinfo
  {journal} {Phys. Rev. X}\ }\textbf {\bibinfo {volume} {12}},\ \bibinfo
  {pages} {011033} (\bibinfo {year} {2022})}\BibitemShut {NoStop}%
\bibitem [{\citenamefont
  {Rao}(2016)}]{https://doi.org/10.48550/arxiv.1603.02821}%
  \BibitemOpen
  \bibfield  {author} {\bibinfo {author} {\bibfnamefont {S.}~\bibnamefont
  {Rao}},\ }\href {https://arxiv.org/abs/1603.02821} {\bibinfo {title} {Weyl
  semi-metals : a short review}} (\bibinfo {year} {2016})\BibitemShut {NoStop}%
\bibitem [{\citenamefont {Colladay}\ and\ \citenamefont
  {Kostelecky}(1998)}]{Colladay:1998fq}%
  \BibitemOpen
  \bibfield  {author} {\bibinfo {author} {\bibfnamefont {D.}~\bibnamefont
  {Colladay}}\ and\ \bibinfo {author} {\bibfnamefont {V.~A.}\ \bibnamefont
  {Kostelecky}},\ }\href {https://doi.org/10.1103/PhysRevD.58.116002}
  {\bibfield  {journal} {\bibinfo  {journal} {Phys. Rev. D}\ }\textbf {\bibinfo
  {volume} {58}},\ \bibinfo {pages} {116002} (\bibinfo {year} {1998})},\
  \Eprint {https://arxiv.org/abs/hep-ph/9809521} {arXiv:hep-ph/9809521}
  \BibitemShut {NoStop}%
\bibitem [{\citenamefont {Zyuzin}\ and\ \citenamefont
  {Burkov}(2011)}]{PhysRevB.83.195413}%
  \BibitemOpen
  \bibfield  {author} {\bibinfo {author} {\bibfnamefont {A.~A.}\ \bibnamefont
  {Zyuzin}}\ and\ \bibinfo {author} {\bibfnamefont {A.~A.}\ \bibnamefont
  {Burkov}},\ }\href {https://link.aps.org/doi/10.1103/PhysRevB.83.195413}
  {\bibfield  {journal} {\bibinfo  {journal} {Phys. Rev. B}\ } (\bibinfo {year}
  {2011})}\BibitemShut {NoStop}%
\bibitem [{\citenamefont {Castro~Neto}\ \emph {et~al.}(2009)\citenamefont
  {Castro~Neto}, \citenamefont {Guinea}, \citenamefont {Peres}, \citenamefont
  {Novoselov},\ and\ \citenamefont {Geim}}]{RevModPhys.81.109}%
  \BibitemOpen
  \bibfield  {author} {\bibinfo {author} {\bibfnamefont {A.~H.}\ \bibnamefont
  {Castro~Neto}}, \bibinfo {author} {\bibfnamefont {F.}~\bibnamefont {Guinea}},
  \bibinfo {author} {\bibfnamefont {N.~M.~R.}\ \bibnamefont {Peres}}, \bibinfo
  {author} {\bibfnamefont {K.~S.}\ \bibnamefont {Novoselov}},\ and\ \bibinfo
  {author} {\bibfnamefont {A.~K.}\ \bibnamefont {Geim}},\ }\href
  {https://link.aps.org/doi/10.1103/RevModPhys.81.109} {\bibfield  {journal}
  {\bibinfo  {journal} {Rev. Mod. Phys.}\ } (\bibinfo {year}
  {2009})}\BibitemShut {NoStop}%
\bibitem [{\citenamefont {Ashby}\ and\ \citenamefont
  {Carbotte}(2013)}]{PhysRevB.87.245131}%
  \BibitemOpen
  \bibfield  {author} {\bibinfo {author} {\bibfnamefont {P.~E.~C.}\
  \bibnamefont {Ashby}}\ and\ \bibinfo {author} {\bibfnamefont {J.~P.}\
  \bibnamefont {Carbotte}},\ }\href
  {https://link.aps.org/doi/10.1103/PhysRevB.87.245131} {\bibfield  {journal}
  {\bibinfo  {journal} {Phys. Rev. B}\ } (\bibinfo {year} {2013})}\BibitemShut
  {NoStop}%
\bibitem [{\citenamefont {Das}\ \emph {et~al.}(2020)\citenamefont {Das},
  \citenamefont {Singh},\ and\ \citenamefont
  {Agarwal}}]{PhysRevResearch.2.033511}%
  \BibitemOpen
  \bibfield  {author} {\bibinfo {author} {\bibfnamefont {K.}~\bibnamefont
  {Das}}, \bibinfo {author} {\bibfnamefont {S.~K.}\ \bibnamefont {Singh}},\
  and\ \bibinfo {author} {\bibfnamefont {A.}~\bibnamefont {Agarwal}},\ }\href
  {https://link.aps.org/doi/10.1103/PhysRevResearch.2.033511} {\bibfield
  {journal} {\bibinfo  {journal} {Phys. Rev. Res.}\ }\textbf {\bibinfo {volume}
  {2}},\ \bibinfo {pages} {033511} (\bibinfo {year} {2020})}\BibitemShut
  {NoStop}%
\bibitem [{\citenamefont {Landsteiner}(2016)}]{landsteiner_notes_2016}%
  \BibitemOpen
  \bibfield  {author} {\bibinfo {author} {\bibfnamefont {K.}~\bibnamefont
  {Landsteiner}},\ }\href {https://doi.org/10.5506/APhysPolB.47.2617}
  {\bibfield  {journal} {\bibinfo  {journal} {Acta Physica Polonica B}\
  }\textbf {\bibinfo {volume} {47}},\ \bibinfo {pages} {2617} (\bibinfo {year}
  {2016})}\BibitemShut {NoStop}%
\bibitem [{\citenamefont {Lu}\ \emph {et~al.}(2015{\natexlab{b}})\citenamefont
  {Lu}, \citenamefont {Zhang},\ and\ \citenamefont
  {Shen}}]{lu_high-field_2015}%
  \BibitemOpen
  \bibfield  {author} {\bibinfo {author} {\bibfnamefont {H.-Z.}\ \bibnamefont
  {Lu}}, \bibinfo {author} {\bibfnamefont {S.-B.}\ \bibnamefont {Zhang}},\ and\
  \bibinfo {author} {\bibfnamefont {S.-Q.}\ \bibnamefont {Shen}},\ }\href
  {https://doi.org/10.1103/PhysRevB.92.045203} {\bibfield  {journal} {\bibinfo
  {journal} {Physical Review B}\ }\textbf {\bibinfo {volume} {92}},\ \bibinfo
  {pages} {045203} (\bibinfo {year} {2015}{\natexlab{b}})}\BibitemShut
  {NoStop}%
\bibitem [{\citenamefont {Zhang}\ \emph {et~al.}(2017)\citenamefont {Zhang},
  \citenamefont {Xu}, \citenamefont {Wang}, \citenamefont {Lin}, \citenamefont
  {Du}, \citenamefont {Guo}, \citenamefont {Lee}, \citenamefont {Lu},
  \citenamefont {Feng}, \citenamefont {Huang}, \citenamefont {Chang},
  \citenamefont {Hsu}, \citenamefont {Liu}, \citenamefont {Lin}, \citenamefont
  {Li}, \citenamefont {Zhang}, \citenamefont {Zhang}, \citenamefont {Xie},
  \citenamefont {Neupert}, \citenamefont {Hasan}, \citenamefont {Lu},
  \citenamefont {Wang},\ and\ \citenamefont
  {Jia}}]{zhang_magnetic-tunnelling-induced_2017}%
  \BibitemOpen
  \bibfield  {author} {\bibinfo {author} {\bibfnamefont {C.-L.}\ \bibnamefont
  {Zhang}}, \bibinfo {author} {\bibfnamefont {S.-Y.}\ \bibnamefont {Xu}},
  \bibinfo {author} {\bibfnamefont {C.~M.}\ \bibnamefont {Wang}}, \bibinfo
  {author} {\bibfnamefont {Z.}~\bibnamefont {Lin}}, \bibinfo {author}
  {\bibfnamefont {Z.~Z.}\ \bibnamefont {Du}}, \bibinfo {author} {\bibfnamefont
  {C.}~\bibnamefont {Guo}}, \bibinfo {author} {\bibfnamefont {C.-C.}\
  \bibnamefont {Lee}}, \bibinfo {author} {\bibfnamefont {H.}~\bibnamefont
  {Lu}}, \bibinfo {author} {\bibfnamefont {Y.}~\bibnamefont {Feng}}, \bibinfo
  {author} {\bibfnamefont {S.-M.}\ \bibnamefont {Huang}}, \bibinfo {author}
  {\bibfnamefont {G.}~\bibnamefont {Chang}}, \bibinfo {author} {\bibfnamefont
  {C.-H.}\ \bibnamefont {Hsu}}, \bibinfo {author} {\bibfnamefont
  {H.}~\bibnamefont {Liu}}, \bibinfo {author} {\bibfnamefont {H.}~\bibnamefont
  {Lin}}, \bibinfo {author} {\bibfnamefont {L.}~\bibnamefont {Li}}, \bibinfo
  {author} {\bibfnamefont {C.}~\bibnamefont {Zhang}}, \bibinfo {author}
  {\bibfnamefont {J.}~\bibnamefont {Zhang}}, \bibinfo {author} {\bibfnamefont
  {X.-C.}\ \bibnamefont {Xie}}, \bibinfo {author} {\bibfnamefont
  {T.}~\bibnamefont {Neupert}}, \bibinfo {author} {\bibfnamefont {M.~Z.}\
  \bibnamefont {Hasan}}, \bibinfo {author} {\bibfnamefont {H.-Z.}\ \bibnamefont
  {Lu}}, \bibinfo {author} {\bibfnamefont {J.}~\bibnamefont {Wang}},\ and\
  \bibinfo {author} {\bibfnamefont {S.}~\bibnamefont {Jia}},\ }\href
  {https://doi.org/10.1038/nphys4183} {\bibfield  {journal} {\bibinfo
  {journal} {Nature Physics}\ }\textbf {\bibinfo {volume} {13}},\ \bibinfo
  {pages} {979} (\bibinfo {year} {2017})},\ \bibinfo {note} {number:
  10}\BibitemShut {NoStop}%
\bibitem [{\citenamefont {Baggioli}\ \emph {et~al.}(2018)\citenamefont
  {Baggioli}, \citenamefont {Padhi}, \citenamefont {Phillips},\ and\
  \citenamefont {Setty}}]{baggioli_conjecture_2018}%
  \BibitemOpen
  \bibfield  {author} {\bibinfo {author} {\bibfnamefont {M.}~\bibnamefont
  {Baggioli}}, \bibinfo {author} {\bibfnamefont {B.}~\bibnamefont {Padhi}},
  \bibinfo {author} {\bibfnamefont {P.~W.}\ \bibnamefont {Phillips}},\ and\
  \bibinfo {author} {\bibfnamefont {C.}~\bibnamefont {Setty}},\ }\href
  {https://doi.org/10.1007/JHEP07(2018)049} {\bibfield  {journal} {\bibinfo
  {journal} {Journal of High Energy Physics}\ }\textbf {\bibinfo {volume}
  {2018}},\ \bibinfo {pages} {49} (\bibinfo {year} {2018})}\BibitemShut
  {NoStop}%
\bibitem [{\citenamefont {Li}\ \emph {et~al.}(2016)\citenamefont {Li},
  \citenamefont {Kharzeev}, \citenamefont {Zhang}, \citenamefont {Huang},
  \citenamefont {Pletikosi{\'{c}}}, \citenamefont {Fedorov}, \citenamefont
  {Zhong}, \citenamefont {Schneeloch}, \citenamefont {Gu},\ and\ \citenamefont
  {Valla}}]{Li_2016}%
  \BibitemOpen
  \bibfield  {author} {\bibinfo {author} {\bibfnamefont {Q.}~\bibnamefont
  {Li}}, \bibinfo {author} {\bibfnamefont {D.~E.}\ \bibnamefont {Kharzeev}},
  \bibinfo {author} {\bibfnamefont {C.}~\bibnamefont {Zhang}}, \bibinfo
  {author} {\bibfnamefont {Y.}~\bibnamefont {Huang}}, \bibinfo {author}
  {\bibfnamefont {I.}~\bibnamefont {Pletikosi{\'{c}}}}, \bibinfo {author}
  {\bibfnamefont {A.~V.}\ \bibnamefont {Fedorov}}, \bibinfo {author}
  {\bibfnamefont {R.~D.}\ \bibnamefont {Zhong}}, \bibinfo {author}
  {\bibfnamefont {J.~A.}\ \bibnamefont {Schneeloch}}, \bibinfo {author}
  {\bibfnamefont {G.~D.}\ \bibnamefont {Gu}},\ and\ \bibinfo {author}
  {\bibfnamefont {T.}~\bibnamefont {Valla}},\ }\href
  {https://doi.org/10.1038%2Fnphys3648} {\bibfield  {journal} {\bibinfo
  {journal} {Nature Physics}\ }\textbf {\bibinfo {volume} {12}},\ \bibinfo
  {pages} {550} (\bibinfo {year} {2016})}\BibitemShut {NoStop}%
\bibitem [{\citenamefont {Zeng}\ \emph {et~al.}(2023)\citenamefont {Zeng},
  \citenamefont {Nandy}, \citenamefont {Liu}, \citenamefont {Tewari},\ and\
  \citenamefont {Yao}}]{Zeng_2023}%
  \BibitemOpen
  \bibfield  {author} {\bibinfo {author} {\bibfnamefont {C.}~\bibnamefont
  {Zeng}}, \bibinfo {author} {\bibfnamefont {S.}~\bibnamefont {Nandy}},
  \bibinfo {author} {\bibfnamefont {P.}~\bibnamefont {Liu}}, \bibinfo {author}
  {\bibfnamefont {S.}~\bibnamefont {Tewari}},\ and\ \bibinfo {author}
  {\bibfnamefont {Y.}~\bibnamefont {Yao}},\ }\href
  {https://doi.org/10.1103%2Fphysrevb.107.l081107} {\bibfield  {journal}
  {\bibinfo  {journal} {Physical Review B}\ }\textbf {\bibinfo {volume} {107}}
  (\bibinfo {year} {2023})}\BibitemShut {NoStop}%
\bibitem [{\citenamefont {Das}\ \emph {et~al.}(2023)\citenamefont {Das},
  \citenamefont {Das},\ and\ \citenamefont {Agarwal}}]{Das:2023onx}%
  \BibitemOpen
  \bibfield  {author} {\bibinfo {author} {\bibfnamefont {S.}~\bibnamefont
  {Das}}, \bibinfo {author} {\bibfnamefont {K.}~\bibnamefont {Das}},\ and\
  \bibinfo {author} {\bibfnamefont {A.}~\bibnamefont {Agarwal}},\ }\href@noop
  {} {\  (\bibinfo {year} {2023})},\ \Eprint {https://arxiv.org/abs/2301.02965}
  {arXiv:2301.02965 [cond-mat.mes-hall]} \BibitemShut {NoStop}%
\bibitem [{\citenamefont {Das}\ and\ \citenamefont
  {Agarwal}(2020)}]{PhysRevResearch.2.013088}%
  \BibitemOpen
  \bibfield  {author} {\bibinfo {author} {\bibfnamefont {K.}~\bibnamefont
  {Das}}\ and\ \bibinfo {author} {\bibfnamefont {A.}~\bibnamefont {Agarwal}},\
  }\href {https://link.aps.org/doi/10.1103/PhysRevResearch.2.013088} {\bibfield
   {journal} {\bibinfo  {journal} {Phys. Rev. Res.}\ }\textbf {\bibinfo
  {volume} {2}},\ \bibinfo {pages} {013088} (\bibinfo {year}
  {2020})}\BibitemShut {NoStop}%
\bibitem [{\citenamefont {Sun}\ and\ \citenamefont {Yang}(2016)}]{Sun:2016gpy}%
  \BibitemOpen
  \bibfield  {author} {\bibinfo {author} {\bibfnamefont {Y.-W.}\ \bibnamefont
  {Sun}}\ and\ \bibinfo {author} {\bibfnamefont {Q.}~\bibnamefont {Yang}},\
  }\href {https://doi.org/10.1007/JHEP09(2016)122} {\bibfield  {journal}
  {\bibinfo  {journal} {JHEP}\ }\textbf {\bibinfo {volume} {09}},\ \bibinfo
  {pages} {122}},\ \Eprint {https://arxiv.org/abs/1603.02624} {arXiv:1603.02624
  [hep-th]} \BibitemShut {NoStop}%
\bibitem [{\citenamefont {Baggioli}\ \emph {et~al.}(2023)\citenamefont
  {Baggioli}, \citenamefont {Liu},\ and\ \citenamefont
  {Wu}}]{Baggioli:2023ynu}%
  \BibitemOpen
  \bibfield  {author} {\bibinfo {author} {\bibfnamefont {M.}~\bibnamefont
  {Baggioli}}, \bibinfo {author} {\bibfnamefont {Y.}~\bibnamefont {Liu}},\ and\
  \bibinfo {author} {\bibfnamefont {X.-M.}\ \bibnamefont {Wu}},\ }\href@noop {}
  {\  (\bibinfo {year} {2023})},\ \Eprint {https://arxiv.org/abs/2302.11096}
  {arXiv:2302.11096 [hep-th]} \BibitemShut {NoStop}%
\bibitem [{\citenamefont {Chernodub}\ \emph {et~al.}(2022)\citenamefont
  {Chernodub}, \citenamefont {Ferreiros}, \citenamefont {Grushin},
  \citenamefont {Landsteiner},\ and\ \citenamefont
  {Vozmediano}}]{Chernodub:2021nff}%
  \BibitemOpen
  \bibfield  {author} {\bibinfo {author} {\bibfnamefont {M.~N.}\ \bibnamefont
  {Chernodub}}, \bibinfo {author} {\bibfnamefont {Y.}~\bibnamefont
  {Ferreiros}}, \bibinfo {author} {\bibfnamefont {A.~G.}\ \bibnamefont
  {Grushin}}, \bibinfo {author} {\bibfnamefont {K.}~\bibnamefont
  {Landsteiner}},\ and\ \bibinfo {author} {\bibfnamefont {M.~A.~H.}\
  \bibnamefont {Vozmediano}},\ }\href
  {https://doi.org/10.1016/j.physrep.2022.06.002} {\bibfield  {journal}
  {\bibinfo  {journal} {Phys. Rept.}\ }\textbf {\bibinfo {volume} {977}},\
  \bibinfo {pages} {1} (\bibinfo {year} {2022})},\ \Eprint
  {https://arxiv.org/abs/2110.05471} {arXiv:2110.05471 [cond-mat.mes-hall]}
  \BibitemShut {NoStop}%
\bibitem [{\citenamefont {Gooth}\ \emph {et~al.}(2017)\citenamefont {Gooth},
  \citenamefont {Niemann}, \citenamefont {Meng}, \citenamefont {Grushin},
  \citenamefont {Landsteiner}, \citenamefont {Gotsmann}, \citenamefont
  {Menges}, \citenamefont {Schmidt}, \citenamefont {Shekhar}, \citenamefont
  {Sü{\ss}}, \citenamefont {Hühne}, \citenamefont {Rellinghaus},
  \citenamefont {Felser}, \citenamefont {Yan},\ and\ \citenamefont
  {Nielsch}}]{Gooth_2017}%
  \BibitemOpen
  \bibfield  {author} {\bibinfo {author} {\bibfnamefont {J.}~\bibnamefont
  {Gooth}}, \bibinfo {author} {\bibfnamefont {A.~C.}\ \bibnamefont {Niemann}},
  \bibinfo {author} {\bibfnamefont {T.}~\bibnamefont {Meng}}, \bibinfo {author}
  {\bibfnamefont {A.~G.}\ \bibnamefont {Grushin}}, \bibinfo {author}
  {\bibfnamefont {K.}~\bibnamefont {Landsteiner}}, \bibinfo {author}
  {\bibfnamefont {B.}~\bibnamefont {Gotsmann}}, \bibinfo {author}
  {\bibfnamefont {F.}~\bibnamefont {Menges}}, \bibinfo {author} {\bibfnamefont
  {M.}~\bibnamefont {Schmidt}}, \bibinfo {author} {\bibfnamefont
  {C.}~\bibnamefont {Shekhar}}, \bibinfo {author} {\bibfnamefont
  {V.}~\bibnamefont {Sü{\ss}}}, \bibinfo {author} {\bibfnamefont
  {R.}~\bibnamefont {Hühne}}, \bibinfo {author} {\bibfnamefont
  {B.}~\bibnamefont {Rellinghaus}}, \bibinfo {author} {\bibfnamefont
  {C.}~\bibnamefont {Felser}}, \bibinfo {author} {\bibfnamefont
  {B.}~\bibnamefont {Yan}},\ and\ \bibinfo {author} {\bibfnamefont
  {K.}~\bibnamefont {Nielsch}},\ }\href {https://doi.org/10.1038/nature23005}
  {\bibfield  {journal} {\bibinfo  {journal} {Nature}\ }\textbf {\bibinfo
  {volume} {547}},\ \bibinfo {pages} {324} (\bibinfo {year}
  {2017})}\BibitemShut {NoStop}%
\bibitem [{\citenamefont {Schindler}\ \emph {et~al.}(2020)\citenamefont
  {Schindler} \emph {et~al.}}]{Schindler:2018wrd}%
  \BibitemOpen
  \bibfield  {author} {\bibinfo {author} {\bibfnamefont {C.}~\bibnamefont
  {Schindler}} \emph {et~al.},\ }\href
  {https://doi.org/10.1103/PhysRevB.101.125119} {\bibfield  {journal} {\bibinfo
   {journal} {Phys. Rev. B}\ }\textbf {\bibinfo {volume} {101}},\ \bibinfo
  {pages} {125119} (\bibinfo {year} {2020})},\ \Eprint
  {https://arxiv.org/abs/1810.02300} {arXiv:1810.02300 [cond-mat.mes-hall]}
  \BibitemShut {NoStop}%
\bibitem [{\citenamefont {Vu}\ \emph {et~al.}(2021)\citenamefont {Vu},
  \citenamefont {Zhang}, \citenamefont {{\c{S}}ahin}, \citenamefont
  {Flatt{\'{e}}}, \citenamefont {Trivedi},\ and\ \citenamefont
  {Heremans}}]{Vu_2021}%
  \BibitemOpen
  \bibfield  {author} {\bibinfo {author} {\bibfnamefont {D.}~\bibnamefont
  {Vu}}, \bibinfo {author} {\bibfnamefont {W.}~\bibnamefont {Zhang}}, \bibinfo
  {author} {\bibfnamefont {C.}~\bibnamefont {{\c{S}}ahin}}, \bibinfo {author}
  {\bibfnamefont {M.~E.}\ \bibnamefont {Flatt{\'{e}}}}, \bibinfo {author}
  {\bibfnamefont {N.}~\bibnamefont {Trivedi}},\ and\ \bibinfo {author}
  {\bibfnamefont {J.~P.}\ \bibnamefont {Heremans}},\ }\href
  {https://doi.org/10.1038/s41563-021-00983-8} {\bibfield  {journal} {\bibinfo
  {journal} {Nature Materials}\ }\textbf {\bibinfo {volume} {20}},\ \bibinfo
  {pages} {1525} (\bibinfo {year} {2021})}\BibitemShut {NoStop}%
\bibitem [{\citenamefont {Andrade}(2017)}]{Andrade:2017jmt}%
  \BibitemOpen
  \bibfield  {author} {\bibinfo {author} {\bibfnamefont {T.}~\bibnamefont
  {Andrade}}\ }(\bibinfo {year} {2017})\ \Eprint
  {https://arxiv.org/abs/1712.00548} {arXiv:1712.00548 [hep-th]} \BibitemShut
  {NoStop}%
\bibitem [{\citenamefont {Baggioli}(2019)}]{Baggioli_2019}%
  \BibitemOpen
  \bibfield  {author} {\bibinfo {author} {\bibfnamefont {M.}~\bibnamefont
  {Baggioli}},\ }\href {https://doi.org/10.1007/978-3-030-35184-7} {\emph
  {\bibinfo {title} {Applied Holography}}}\ (\bibinfo  {publisher} {Springer
  International Publishing},\ \bibinfo {year} {2019})\BibitemShut {NoStop}%
\end{thebibliography}%

\end{document}